\documentclass{article} 
\usepackage{iclr2027_conference,times}

\usepackage{amsmath,amsfonts,bm}

\def\eqref#1{equation~\ref{#1}}

\def\1{\bm{1}}

\DeclareMathAlphabet{\mathsfit}{\encodingdefault}{\sfdefault}{m}{sl}
\SetMathAlphabet{\mathsfit}{bold}{\encodingdefault}{\sfdefault}{bx}{n}

\usepackage{hyperref}
\usepackage{url}
\usepackage{booktabs}
\usepackage{multirow}
\usepackage{makecell}
\usepackage[table]{xcolor}
\usepackage{pifont}
\usepackage{graphicx}
\usepackage{diagbox}
\usepackage{adjustbox}
\usepackage{subcaption}
\usepackage{tabularx}
\usepackage{array}
\usepackage{adjustbox}
\usepackage{subcaption}
\usepackage{caption}
\usepackage{enumitem}
\usepackage{placeins}
\usepackage{tablefootnote}

\newcommand{\cmark}{\textcolor{green!55!black}{\ding{51}}}
\newcommand{\xmark}{\textcolor{red!70!black}{\ding{55}}}
\newcommand{\pmark}{\textcolor{orange!85!black}{\ding{108}}}

\newcommand{\amark}{\cmark\textsuperscript{\tiny \textbf{A}}}

\newcommand{\avmark}{\cmark\textsuperscript{\tiny \textbf{AV}}}
\newcommand{\asepmark}{\cmark\textsuperscript{\tiny \textbf{A/V}}}

\title{SAVU-BENCH: A Real-World Benchmark for Spatial Audio-Visual Understanding}

\author{Yu Chen$^1$, Ruihang Liu$^2$, Yangguang Xu$^2$, Xinyue Jiang$^2$, \\
{\bfseries Mohammed Bennamoun$^1$, Farid Boussaid$^1$, Xinyuan Qian$^2$, Qiuhong Ke$^3$}\\[0.35em]
$^1$University of Western Australia\\
$^2$University of Science and Technology Beijing\\
$^3$Monash University
}

\iclrfinalcopy 
\begin{document}

\maketitle

\vspace{-0.4em}

\begin{abstract}
Spatial audio-visual understanding requires models to recognize not only what is present, but also where events occur and how they relate across modalities. Existing benchmarks often rely on simulated scenes, evaluate isolated spatial skills, and provide limited diagnostic insight into failure modes. We introduce \textbf{SAVU-Bench}, a real-world benchmark that systematically evaluates spatial audio-visual understanding across three capability levels and seven evaluation tasks. We further introduce \textbf{SAVU-Diag}, a scene-linked diagnostic set that decomposes reasoning questions into their prerequisite grounding and alignment sub-tasks. Evaluation of 12 representative models on SAVU-Bench reveals that while visual spatial grounding is relatively mature, spatial perception involving audio remains a primary bottleneck. SAVU-Diag further demonstrates that most reasoning errors co-occur with failures on these prerequisite tasks, though reasoning gaps persist even when prerequisites are correctly resolved. Motivated by these findings, we introduce \textbf{SAVU-EA}, a training-free evidence-augmented baseline that makes spatial cues more explicit. While SAVU-EA substantially improves spatial grounding and joint matching, high-level spatial reasoning remains challenging. Our findings highlight the urgent need for both robust spatial audio perception and deeper integration of cross-modal spatial relations.
\end{abstract}

\begin{center}
    \includegraphics[
        width=0.95\textwidth
    ]{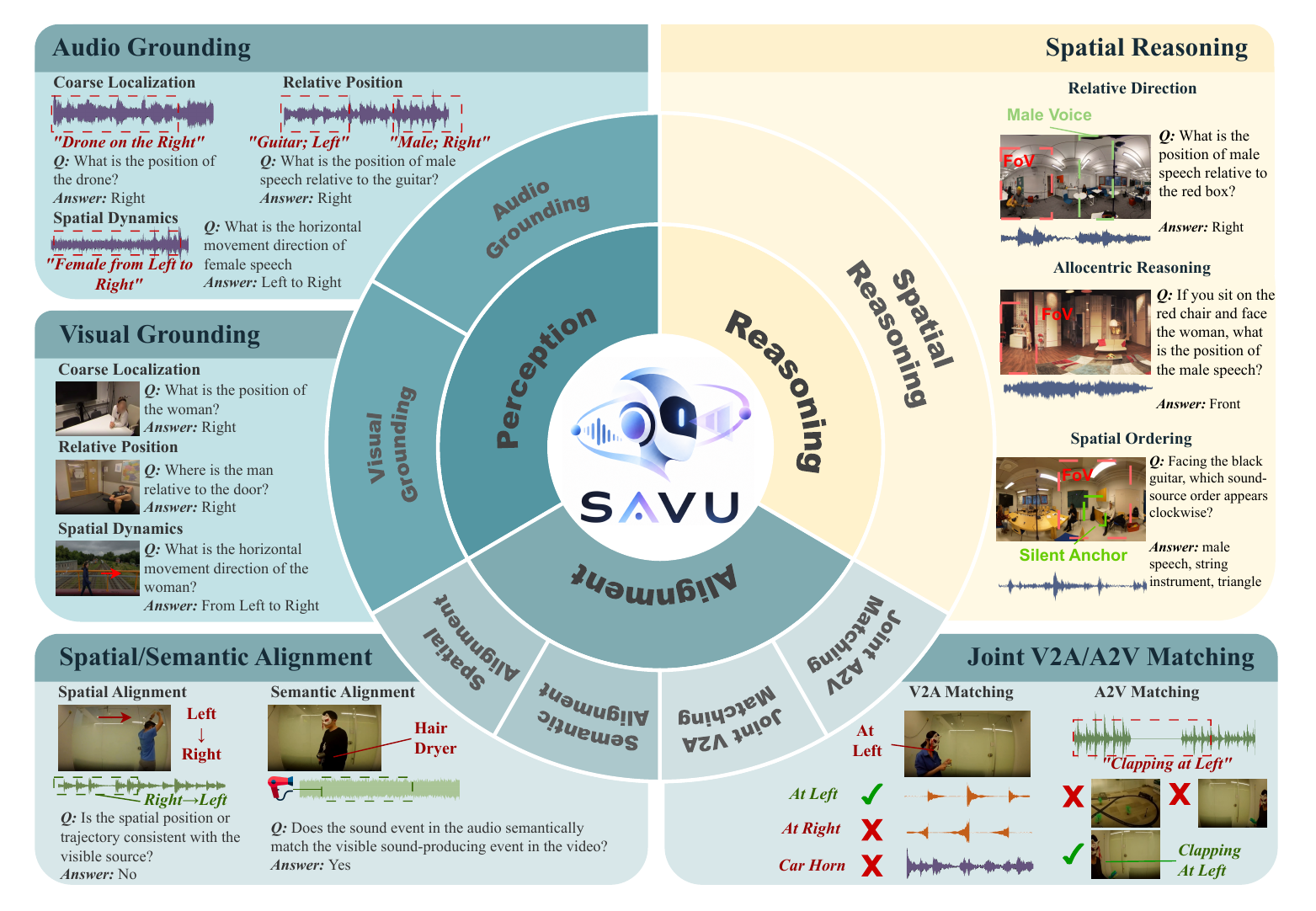}

    \captionof{figure}{
        \textbf{Overview of SAVU-Bench.} SAVU-Bench evaluates spatial audio-visual understanding across three capability levels: \emph{Perception}, \emph{Alignment}, and \emph{Reasoning}. Representative examples from seven evaluation tasks illustrate the progression from modality-specific spatial grounding to cross-modal alignment and spatial reasoning. FoV denotes field of view.
    }
    \label{fig:overview}
\end{center}

\section{Introduction}

Recent advances in Audio-Visual Large Language Models (AV-LLMs) have driven substantial progress in multimodal scene understanding by jointly processing visual and acoustic information~\citep{qwen25,watchlisten,minicpm}. However, recognizing what is seen and heard addresses only part of scene understanding. Models must also localize visual entities and acoustic sources while resolving their cross-modal spatial relationships~\citep{avssan,spatialomni,ryu2026hear}. We refer to this capability as \emph{spatial audio-visual understanding}, which unifies semantic recognition, spatial perception, and cross-modal reasoning.

Existing benchmarks have begun to evaluate spatial capabilities through source localization~\citep{starbench,spatialomni}, cross-modal matching~\citep{xmodbench}, spatial question answering~\citep{jaeger}, and reasoning~\citep{savvy}. However, three key limitations remain. First, many spatial benchmarks rely heavily on simulated environments~\citep{jaeger,spatialomni,starbench}. Simulation provides accurate spatial annotations and controlled configurations, but does not fully capture the complexity of real-world recordings. Evaluation on real scenes is therefore important for understanding how well current models generalize to realistic spatial conditions. Second, existing benchmarks often evaluate spatial skills in isolation. This makes it difficult to obtain a systematic view of how perception, cross-modal alignment, and reasoning develop within the same evaluation setting. Third, task-level accuracy metrics lack diagnostic visibility into failure mechanisms; specifically, reasoning errors cannot be disentangled from underlying failures in spatial localization, cross-modal alignment, or relational composition.

To address these limitations, we introduce the \textbf{S}patial \textbf{A}udio-\textbf{V}isual \textbf{U}nderstanding benchmark, \textbf{SAVU-Bench}. SAVU-Bench is built from real-world spatial audio-visual recordings and systematically evaluates three capability levels: Perception, Alignment, and Reasoning. As illustrated in Figure~\ref{fig:overview}, these capabilities are instantiated across seven evaluation tasks that cover progressively more demanding forms of spatial understanding. The benchmark is constructed from six real-world data sources through a Human-LLM annotation pipeline with expert verification. Rather than evaluating individual spatial skills in isolation, SAVU-Bench provides a unified framework for evaluating how current models perceive, align, and reason over cross-modal spatial information.

However, a benchmark-level score alone cannot reveal why spatial reasoning fails. We therefore introduce \textbf{SAVU-Diag}, a scene-linked diagnostic set designed to connect reasoning outcomes with the prerequisite capabilities involved in solving each question. For every reasoning instance, SAVU-Diag provides corresponding grounding and alignment tasks derived from the same scene and evaluates them independently. This structure enables us to disentangle whether reasoning errors co-occur with prerequisite failures or persist despite correct prerequisite resolution, providing fine-grained behavioral diagnosis beyond monolithic accuracy metrics.

We evaluate 12 representative audio-visual models on SAVU-Bench. The results reveal an imbalance across spatial capabilities. Current models perform well on visual grounding, while spatial understanding involving audio remains substantially weaker. Performance further drops when models must combine semantic and spatial relations or use them for spatial reasoning. SAVU-Diag offers granular diagnostic insight into these failures. Models with similar reasoning accuracy can exhibit different grounding and alignment error patterns, and some reasoning failures persist even when all evaluated prerequisite tasks are answered correctly. These observations suggest two distinct error modes: spatial evidence acquisition (perceptual and alignment failures) versus spatial evidence composition (reasoning failures over valid premises).

Motivated by these observations, we further introduce \textbf{SAVU-EA}, a training-free baseline that augments an AV-LLM with explicit spatial cues extracted from the audio and visual inputs. While SAVU-EA substantially improves auditory grounding and joint cross-modal matching, spatial reasoning remains challenging. This result provides an informative intervention: while stronger spatial perception boosts downstream performance, reliable spatial reasoning requires more than better perceptual cues alone.

Our contributions are summarized as follows:
\begin{itemize}[leftmargin=1.3em, itemsep=0.15em, topsep=0.25em, parsep=0pt, partopsep=0pt]

\item We introduce \textbf{SAVU-Bench}, a real-world benchmark curated from six diverse data sources that systematically evaluates spatial audio-visual understanding across three capability levels (Perception, Alignment, and Reasoning) and seven distinct evaluation tasks.

\item We introduce \textbf{SAVU-Diag}, a scene-linked diagnostic set that connects reasoning questions with their prerequisite grounding and alignment tasks, enabling fine-grained analysis beyond aggregate task scores.

\item We evaluate 12 representative audio-visual models, uncovering severe bottlenecks in auditory spatial grounding and cross-modal reasoning. Crucially, SAVU-Diag further shows that similar reasoning accuracy can arise from different prerequisite-error patterns.

\item We introduce \textbf{SAVU-EA}, a training-free baseline showing that explicit spatial cues substantially improve auditory grounding and joint matching, while the remaining reasoning gap highlights the need for better composition of cross-modal spatial relations.

\end{itemize}

\section{Related Work}
Broad audio-visual benchmarks have incorporated sound-source grounding and spatial understanding into more general multimodal evaluation \citep{avodyssey,omnibench,maverix,jointavbench}. More specialized benchmarks target individual spatial capabilities, including STAR-Bench \citep{starbench} and SO-Bench \citep{spatialomni} for spatial audio, and AudioCOCO \citep{audiococo}, SpatialSceneQA \citep{jaeger}, XModBench \citep{xmodbench}, and SAVVY-Bench \citep{savvy} for audio-visual localization, matching, or reasoning.

These benchmarks differ in both data source and evaluation scope. Several spatial benchmarks rely largely on simulated environments, which provide controlled spatial annotations but may not reflect the complexity of real recordings. Recent work has also reported performance degradation when transferring models from simulated environments to real-world spatial scenes \citep{jaeger}. Existing real-world benchmarks improve realism, but often focus on a narrower subset of spatial capabilities. In addition, prior evaluations commonly emphasize localization, cross-modal matching, or reasoning separately, making it difficult to assess how these capabilities relate within a unified setting.

SAVU-Bench complements these efforts by combining diverse real-world spatial audio-visual recordings with a unified evaluation of Perception, Alignment, and Reasoning. Beyond task-level evaluation, SAVU-Diag further links reasoning questions with prerequisite grounding and alignment tasks from the same scenes, enabling scene-linked analysis of reasoning failures. Table~\ref{tab:benchmark_comparison} summarizes the comparison with representative benchmarks\footnote{
We categorize SpatialSceneQA's multi-speaker matching tasks as Alignment because they evaluate direct audio--visual source correspondence; in our taxonomy, Reasoning requires spatial relation composition beyond direct source matching.
}.

\begin{table*}[t]
\centering
\footnotesize
\setlength{\tabcolsep}{6.0pt}
\renewcommand{\arraystretch}{1.15}

\caption{
Comparison with representative spatial audio and audio-visual benchmarks.
\cmark, \xmark, and \pmark\ denote full, absent, and partial coverage, respectively.
Superscripts A, AV, and A/V indicate audio-only, joint audio-visual, and separate audio/visual settings.
\textit{Real Rec.} denotes real-recorded data, and \textit{Linked Diag.} denotes scene-linked diagnostic evaluation.
}
\label{tab:benchmark_comparison}

\begin{tabular}{lccccccc}
\toprule
\multirow{2}{*}{\textbf{Benchmark}}
& \multicolumn{3}{c}{\textbf{Basic Setting}}
& \multicolumn{3}{c}{\textbf{Task Scope}}
& \multicolumn{1}{c}{\textbf{Analysis}} \\
\cmidrule(lr){2-4}
\cmidrule(lr){5-7}
\cmidrule(lr){8-8}

&
\textbf{Audio} &
\textbf{Vision} &
\makecell{\textbf{Real}\\\textbf{Rec.}} &
\textbf{Perception} &
\textbf{Alignment} &
\textbf{Reasoning} &
\makecell{\textbf{Linked}\\\textbf{Diag.}}
\\
\midrule

STAR-Bench
& \cmark & \xmark & \pmark
& \amark & \xmark & \amark
& \xmark \\

SO-Bench
& \cmark & \xmark & \pmark
& \amark & \xmark & \amark
& \xmark \\

AudioCOCO
& \cmark & \cmark & \xmark
& \avmark & \pmark
& \xmark
& \xmark \\

SpatialSceneQA
& \cmark & \cmark & \xmark
& \asepmark & \avmark & \xmark
& \xmark \\

SAVVY-Bench
& \cmark & \cmark & \cmark
& \xmark & \xmark & \avmark
& \xmark \\

XModBench
& \cmark & \cmark & \pmark
& \xmark & \avmark & \xmark
& \xmark \\

\midrule
\rowcolor{gray!10}
\textbf{SAVU-Bench}
& \cmark & \cmark & \cmark
& \asepmark & \avmark & \avmark
& \cmark \\

\bottomrule
\end{tabular}

\vspace{2pt}

\end{table*}

\section{SAVU-Bench}

We introduce SAVU-Bench for evaluating spatial audio-visual understanding in real-world scenes across Perception, Alignment, and Reasoning, instantiated through seven evaluation tasks, as shown in Fig.~\ref{fig:overview}. All visual inputs use a limited field of view (FoV), while SAVU-Diag provides scene-linked diagnosis of reasoning failures. The benchmark is built from six real-world sources with human annotation and expert verification.

\subsection{Task Taxonomy}
\label{sec:taxonomy}

\paragraph{Perception.}
Audio Grounding and Visual Grounding evaluate spatial understanding using only the corresponding modality. Both contain three question types: Coarse Localization, Relative Position, and Spatial Dynamics, which respectively assess target location, relations between entities, and changes in position over time. Visual Spatial Dynamics covers both horizontal and depth motion. In all cases, semantic cues identify the queried entity before its spatial state is determined.

\paragraph{Alignment.}
Alignment comprises four evaluation tasks: Semantic Alignment, Spatial Alignment, Audio-to-Video (A2V) Joint Matching, and Video-to-Audio (V2A) Joint Matching. Semantic Alignment tests whether an audible event matches the visible object or activity, while Spatial Alignment tests whether the trajectory of a sound source is consistent with the motion of its visual counterpart. To isolate spatial correspondence from semantic mismatch, spatial negatives preserve category-level semantic compatibility while altering the underlying spatial relation. Joint Matching requires both forms of correspondence: A2V pairs an audio query with three candidate videos, whereas V2A reverses the modalities. Each candidate set contains a match, a semantic negative, and a spatial negative.


\paragraph{Reasoning.} 
Spatial Reasoning evaluates cross-modal relation composition through three question types. Relative Direction locates an out-of-view sound source relative to a visual reference. Spatial Ordering determines the sequence of sound-source directions encountered while rotating in place from a visually specified starting direction. Allocentric Reasoning reinterprets source locations from a virtual observer position and facing direction defined by scene entities.

To reduce direct single-modality shortcuts, reasoning instances are constructed around complementary spatial cues from the two modalities. A typical instance contains a silent visual anchor, an in-view sounding source, and an out-of-view sound source. For the example shown in Fig~\ref{fig:overview}, the visual input may show a red box and a guitarist, while a male speaker is only observable through the audio. To determine the speaker's position relative to the red box, the model must relate the box to the guitarist visually and the guitarist to the speaker acoustically. The in-view sounding source therefore serves as a common reference that connects the visual and acoustic spatial layouts.

Detailed spatial reference construction and task examples are provided in Appendix~\ref{app:spatial_reference} and Appendix~\ref{app:task_details}.

\subsection{SAVU-Diag: Scene-Linked Diagnosis}
\label{sec:diag}

SAVU-Bench provides a broad capability evaluation, but an incorrect reasoning answer alone does not reveal where the failure occurs. We therefore introduce SAVU-Diag to connect reasoning outcomes with the prerequisite tasks needed to solve each question. For every selected reasoning instance, we identify the relevant grounding facts and cross-modal correspondences and convert them into independent grounding and alignment tasks derived from the same scene.

All diagnostic tasks share the same underlying recording, time window, target identities, and reference-frame conventions. Grounding tasks receive only the corresponding modality, whereas alignment and reasoning tasks receive both audio and vision. We can therefore examine whether reasoning errors co-occur with failures in grounding or alignment, or persist even when all evaluated prerequisite tasks are answered correctly. Detailed SAVU-Diag construction is provided in Appendix~\ref{app:savu_diag_construction}.

\subsection{Data Curation Pipeline}
\label{sec:data}

\begin{figure*}[t]
    \centering

    \begin{minipage}[c]{0.72\textwidth}
        \centering
        \includegraphics[width=\linewidth]
        {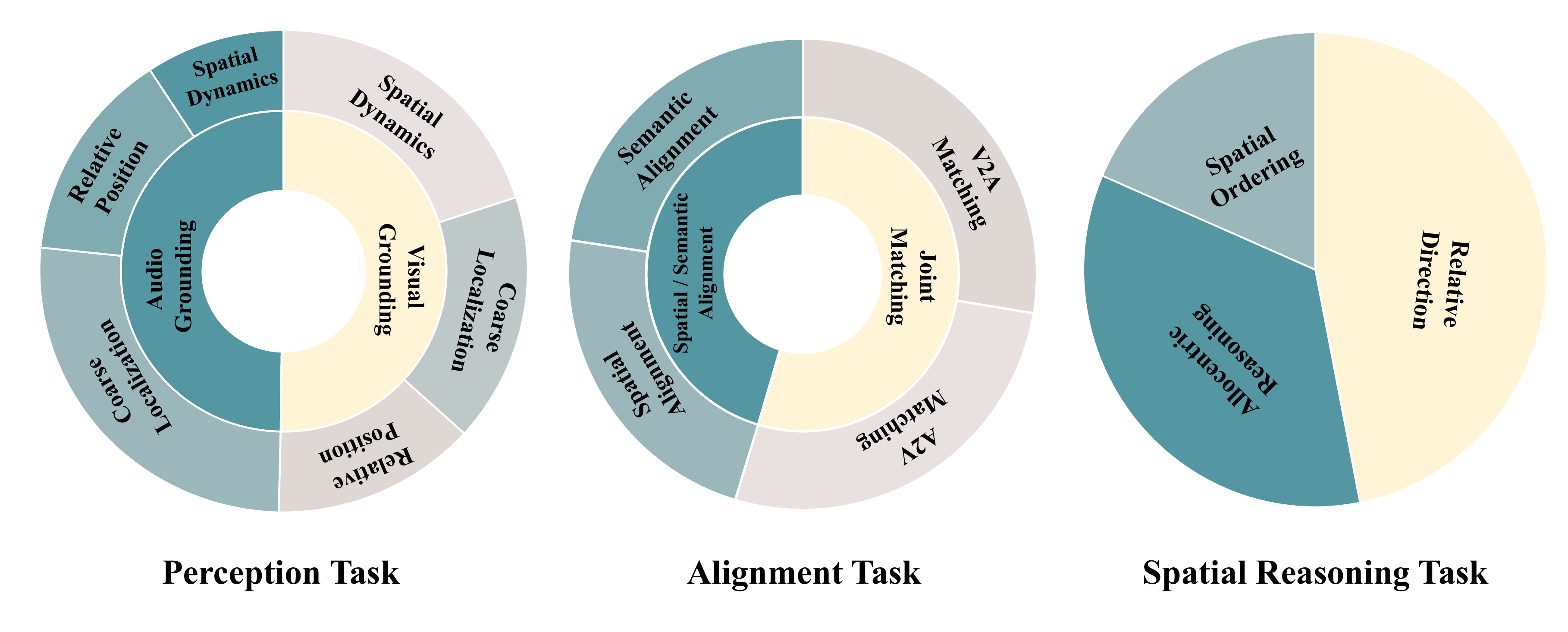}
    \end{minipage}
    \hfill
    \begin{minipage}[c]{0.27\textwidth}
        \centering
        \scriptsize

        \begin{adjustbox}{max width=\linewidth}
        \begin{tabular}{
            @{}
            l
            @{\hspace{0.8em}}
            r
            @{}
        }
            \toprule

            \multicolumn{2}{l}{\textbf{SAVU-Bench}} \\
            \midrule
            Total QAs            & 1,066 \\
            Evaluation Tasks          & 7 \\
            Avg. Duration        & 6.12 s \\
            Avg. Ques. Length & 15.41 words \\

            \midrule

            \multicolumn{2}{l}{\textbf{SAVU-Diag}} \\
            \midrule
            Diagnostic QAs       & 611 \\
            Scenes               & 81 \\
            Avg. QAs / Scene     & 7.54 \\

            \bottomrule
        \end{tabular}
        \end{adjustbox}

    \end{minipage}

    \caption{Distribution of SAVU-Bench tasks and summary statistics for SAVU-Bench and SAVU-Diag.
    }
    \label{fig:savu_distribution_statistics}

\end{figure*}

SAVU-Bench is constructed from six real-world spatial audio-visual sources: MRSAudio \citep{mrsaudio}, STARSS23 \citep{starss23}, YT360 \citep{yt360}, D-SAV360 \citep{dsav360}, MULTIVOX \citep{multivox}, and QoEVAVE \citep{qoevave}. We use a four-stage pipeline with human review of all retained question--answer pairs.


\paragraph{1) Candidate Screening and Scene Structuring.}
We filter invalid or overly short segments and those with limited audio activity, and split long recordings into candidate windows. We construct scene metadata from available annotations, candidate sound-event labels from BEATs \citep{beats} and visual objects and trajectories from Grounded SAM 2 \citep{gm1, gm3}. When spatial annotations are unavailable, OpenVocabularySELD \citep{SELD} estimates sound activity and direction. For panoramic recordings, we extract limited-FoV views and align the corresponding first-order Ambisonics (FOA) audio with the visual reference frame following ViSAGe \citep{kim2025visage}, with more details provided in Appendix~\ref{app:spatial_reference}.


\paragraph{2) Question Candidate Generation.}
We define task-specific templates according to the taxonomy in Sec.~\ref{sec:taxonomy}. For each candidate scene, the structured metadata and templates are provided to GPT-5.4 \citep{gpt54} to identify suitable task types and generate initial QA candidates for Perception and Alignment. These candidates are treated as drafts rather than final annotations and are subsequently verified and refined by human annotators.

\paragraph{3) Human Annotation and Reasoning Construction.}
Annotators remove candidates with ambiguous targets, unclear spatial relations, or insufficient evidence, and refine valid questions by adjusting visual views and audio orientations. Reasoning questions are constructed manually from verified scene relations. For each candidate scene, annotators first identify the visual anchors, sounding sources, and spatial references needed to define the question and its answer. These annotated relations are then provided to GPT-5.4 to generate an initial reasoning chain, which is reviewed and revised by annotators to ensure that every reasoning step is supported by the scene.

SAVU-Diag is derived from these verified reasoning instances. For each reasoning question, we provide its reviewed reasoning chain to GPT-5.4 to identify the grounding facts and cross-modal correspondences required to reach the answer and to organize them into an initial prerequisite chain. Human annotators then refine this chain by removing redundant steps, correcting unsupported relations, and adding any missing prerequisite information. Each retained prerequisite is converted into an independent diagnostic task. Grounding tasks test the required semantic or spatial facts within the corresponding modality, while alignment tasks test the cross-modal correspondences needed to connect these facts. All diagnostic tasks preserve the same scene, time window, target identities, and reference-frame conventions as the original reasoning question. Their answers are verified independently.


\begin{table*}[t]
\centering
\begingroup

\setlength{\tabcolsep}{4.2pt}
\renewcommand{\arraystretch}{1.14}
\footnotesize

\caption{
Results on SAVU-Bench in Average Accuracy (AA, \%).
Best performance in each task is shown in bold.
$^\dagger$ Chance is computed under the same cyclic evaluation protocol.
The detailed calculation is provided in Appendix~\ref{app:chance_calculation}.
}
\label{tab:savu-main-results}

\begin{adjustbox}{max width=\linewidth}
\begin{tabular}{lccccccc}
\toprule

\multirow{2}{*}{\textbf{Model}}
& \multicolumn{2}{c}{\textbf{Perception}}
& \multicolumn{4}{c}{\textbf{Alignment}}
& \multicolumn{1}{c}{\textbf{Reasoning}} \\

\cmidrule(lr){2-3}
\cmidrule(lr){4-7}
\cmidrule(lr){8-8}

&
\shortstack{\textbf{Audio}\\\textbf{Grounding}}
&
\shortstack{\textbf{Visual}\\\textbf{Grounding}}
&
\shortstack{\textbf{Semantic}\\\textbf{Alignment}}
&
\shortstack{\textbf{Spatial}\\\textbf{Alignment}}
&
\shortstack{\textbf{Joint}\\\textbf{A2V}}
&
\shortstack{\textbf{Joint}\\\textbf{V2A}}
&
\shortstack{\textbf{Spatial}\\\textbf{Reasoning}}
\\

\midrule

Chance$^\dagger$
& 50.0
& 50.0
& 50.0
& 50.0
& 33.3
& 33.3
& 42.2 \\

\midrule

\rowcolor{gray!8}
\multicolumn{8}{l}{\textbf{Closed-source Models}} \\

Qwen3.5-Omni-Plus
& 51.2
& 94.8
& 59.7
& 48.0
& 35.2
& 35.3
& 43.8 \\

Qwen3.5-Omni-Flash
& 48.8
& 89.9
& 54.0
& 56.3
& 32.6
& 33.3
& 43.2 \\

Gemini-3.6-Flash
& 52.5
& 93.6
& \textbf{65.7}
& 52.0
& 35.4
& \textbf{42.8}
& 44.8 \\

Gemini-2.5-Pro
& 52.2
& 91.7
& 59.7
& 49.3
& 28.5
& 41.7
& \textbf{45.8} \\

Doubao-Seed-2.0-Lite
& 48.4
& 88.3
& 62.0
& 52.7
& 29.2
& 39.7
& 41.7 \\

\midrule

\rowcolor{gray!8}
\multicolumn{8}{l}{\textbf{Open-weight Models}} \\

Nemotron-3-Nano-Omni
& 53.4
& 94.5
& 50.3
& 51.0
& 22.7
& 29.5
& 44.3 \\

DOTS-3-Note-Preview
& 47.2
& \textbf{95.4}
& 53.0
& 50.3
& 35.4
& 33.2
& 43.2 \\

Qwen2.5-Omni-7B
& 48.1
& 78.5
& 51.7
& 50.0
& 33.1
& 35.7
& 44.3 \\

Spatial-Omni SO-7B
& \textbf{55.6}
& 71.5
& 50.7
& \textbf{58.0}
& 33.3
& 33.0
& \textbf{45.8} \\

MiniCPM-o-4.5
& 48.4
& 84.7
& 58.3
& 49.0
& 32.8
& 33.5
& 43.8 \\

video-SALMONN2+ 7B
& 48.8
& 81.0
& 50.7
& 52.3
& \textbf{35.6}
& 33.5
& 39.6 \\

\shortstack[l]{OmniVideo-7B\\(VITA1.5)}
& 50.3
& 73.9
& 50.3
& 50.0
& 31.6
& 33.0
& 40.6 \\

\bottomrule
\end{tabular}
\end{adjustbox}

\endgroup
\end{table*}

\paragraph{4) Expert Verification.}
Following the expert-based validation procedure in STAR-Bench \citep{starbench}, three domain experts independently answer each question; only those answered correctly by at least two experts are retained. The final datasets contain 1,066 SAVU-Bench questions and 611 SAVU-Diag questions. Figure~\ref{fig:savu_distribution_statistics} summarizes the resulting data distribution and key statistics. Further details on human annotation and expert verification are provided in Appendix~\ref{app:human_annotation}. 

\section{Experiments}

\subsection{Experimental Setup}

\paragraph{Benchmarking Models.}
We evaluate 12 representative audio-visual models, including seven open-weight models and five closed-source models. The evaluation covers general-purpose audio-visual and omni models, including Qwen2.5-Omni-7B \citep{qwen25}, MiniCPM-o-4.5 \citep{minicpm}, video-SALMONN2+ 7B \citep{videosalmonn}, OmniVideo-7B (VITA1.5) \citep{omnivideo}, Nemotron-3-Nano-Omni \citep{nvidia}, DOTS-3-Note-Preview \citep{dots3}, Qwen3.5-Omni-Plus \citep{qwen35}, Qwen3.5-Omni-Flash \citep{qwen35}, Gemini-3.6-Flash \citep{gemini36}, Gemini-2.5-Pro \citep{gemini25}, and Doubao-Seed-2.0-Lite \citep{seed20}. We additionally include Spatial-Omni (SO-7B) \citep{spatialomni}, which is specifically designed for spatial audio understanding.

\paragraph{Evaluation Protocol.}
All SAVU-Bench questions are evaluated in a multiple-choice format. 
For the main benchmark results, we reduce sensitivity to option position by evaluating each question under all cyclic permutations of its answer choices, ensuring that every candidate appears once at each position.
Let $K_i$ denote the number of answer choices for question $i$, and let $c_{ij}\in\{0,1\}$ indicate whether the model answers question $i$ correctly under the $j$-th cyclic ordering. We compute Average Accuracy (AA) over all question--ordering pairs as
\begin{equation}
\mathrm{AA}_{\mathrm{circ}}
=
\frac{
\sum_{i=1}^{N}\sum_{j=1}^{K_i} c_{ij}
}{
\sum_{i=1}^{N} K_i
}
\times 100\%.
\label{eq:circular_aa}
\end{equation}
The correct answer label is updated accordingly under each permutation.
For SAVU-Diag and the human-evaluation subset, each model is evaluated once per question without option permutation. This matches the single-response protocol used for human evaluation and enables direct comparison between model and human performance on the shared subset.


Following SAVVY~\citep{savvy}, we evaluate models through their supported audio interfaces. Spatial-Omni retains the original four-channel FOA input for its dedicated spatial encoder, whereas the other models receive a mono downmix, submitted through their respective local pipelines or APIs. For human evaluation, the same underlying FOA recordings are rendered into binaural audio using head-related transfer functions (HRTFs) \citep{hrtf}. Thus, models and humans are evaluated on the same scenes and questions, with audio representations adapted to their respective interfaces.

We also conduct a human evaluation on a shared subset of the benchmark. We randomly sample 30 questions from each evaluation task to evaluate human performance. Further details on the human evaluation protocol are provided in Appendix~\ref{app:human_annotation}.

\begin{table*}[t]
\centering
\begingroup

\setlength{\tabcolsep}{4.2pt}
\renewcommand{\arraystretch}{1.10}
\footnotesize

\caption{
\textbf{Human comparison on the SAVU-Bench subset.}
Parentheses show the human--model gap.
}
\label{tab:human-comparison}

\begin{adjustbox}{max width=\linewidth}
\begin{tabular}{lccccccc}
\toprule

\textbf{Model}
& \shortstack{\textbf{Audio}\\\textbf{Grounding}}
& \shortstack{\textbf{Visual}\\\textbf{Grounding}}
& \shortstack{\textbf{Semantic}\\\textbf{Alignment}}
& \shortstack{\textbf{Spatial}\\\textbf{Alignment}}
& \shortstack{\textbf{Joint}\\\textbf{A2V}}
& \shortstack{\textbf{Joint}\\\textbf{V2A}}
& \shortstack{\textbf{Spatial}\\\textbf{Reasoning}}
\\

\midrule

Spatial-Omni SO-7B
& \textbf{66.67}
& 66.67
& 53.33
& 50.00
& \textbf{36.67}
& 33.33
& \textbf{46.67} \\

MiniCPM-o-4.5
& 56.67
& 86.67
& 53.33
& \textbf{53.33}
& 26.67
& 20.00
& 43.33 \\

Gemini-3.6-Flash
& 56.67
& 96.67
& 70.00
& 43.33
& \textbf{36.67}
& \textbf{43.33}
& 33.33 \\

Qwen3.5-Omni-Plus
& \textbf{66.67}
& \textbf{100.00}
& \textbf{73.33}
& 46.67
& 33.33
& 20.00
& 40.00 \\

\midrule

\rowcolor{gray!5}
\textbf{Human}
& 89.33 {\scriptsize\textcolor{green!55!black}{(+22.6)}}
& 93.33 {\scriptsize\textcolor{red!70!black}{($-6.67$)}}
& 86.67 {\scriptsize\textcolor{green!55!black}{(+13.33)}}
& 71.67 {\scriptsize\textcolor{green!55!black}{(+18.33)}}
& 66.67 {\scriptsize\textcolor{green!55!black}{(+30.00)}}
& 65.33 {\scriptsize\textcolor{green!55!black}{(+22.00)}}
& 85.00 {\scriptsize\textcolor{green!55!black}{(+38.33)}} \\

\bottomrule
\end{tabular}
\end{adjustbox}

\endgroup
\end{table*}

\subsection{Main Result Analysis}
Table~\ref{tab:savu-main-results} reports the overall results on SAVU-Bench, while Table~\ref{tab:human-comparison} compares representative models with human performance on the shared subset. Additional human-subset results and complementary diagnostic and modality-ablation analyses are provided in Appendix~\ref{sec:complete_results}. Based on Table~\ref{tab:savu-main-results} and Table~\ref{tab:human-comparison}, we make three main observations. 


First, auditory spatial grounding remains limited across all evaluated models. Models perform substantially better on Visual Grounding than on Audio Grounding. Most general-purpose models do not natively support spatial audio and instead process mono audio, which removes the multichannel cues required for spatial localization. This creates a clear system-level bottleneck for auditory spatial perception. Notably, Spatial-Omni directly processes FOA audio through a dedicated spatial encoder, yet achieves only 55.6\% on Audio Grounding. This indicates that native spatial-audio support alone is not sufficient for robust localization in real-world scenes. Together, these results highlight limitations in both spatial-audio support and the real-world generalization of current spatially specialized models.

Second, better performance on individual alignment tasks does not translate into reliable joint matching. Most general-purpose models perform better on Semantic Alignment than on Spatial Alignment, whereas Spatial-Omni shows the opposite pattern. This indicates that semantic and spatial correspondence do not necessarily improve together. The gap becomes more apparent in Joint Matching, where the best A2V and V2A accuracies reach only 35.6\% and 42.8\%, compared with a 33.3\% chance level. Current models therefore struggle to preserve semantic and spatial consistency simultaneously in cross-modal decisions. 

Third, Spatial Reasoning remains challenging. Gemini-2.5-Pro and Spatial-Omni achieve the highest accuracy among the API-based and open-weight models, respectively, both reaching 45.8\%, only 3.6 percentage points above the chance level. Human evaluation in Table~\ref{tab:human-comparison} further highlights the remaining gaps. While the best model slightly exceeds human performance on Visual Grounding, humans retain substantial advantages on other tasks. These results suggest that the main limitations lie in acquiring auditory and cross-modal spatial information and reliably combining these relations for reasoning, rather than in visual spatial grounding alone.

\subsection{Diagnostic Analysis with SAVU-Diag}
SAVU-Bench shows that Spatial Reasoning remains challenging, but aggregate accuracy alone cannot reveal which prerequisite capabilities are associated with these failures. We therefore use SAVU-Diag to analyze reasoning outcomes together with their corresponding grounding and alignment tasks. We report four representative models in the main text, with complete SAVU-Diag and oracle experiment results provided in Appendix~\ref{app:diag_results}.

\begin{figure*}[t]
    \centering

    \begin{subfigure}[t]{0.82\textwidth}
        \centering
        \includegraphics[width=\linewidth]
        {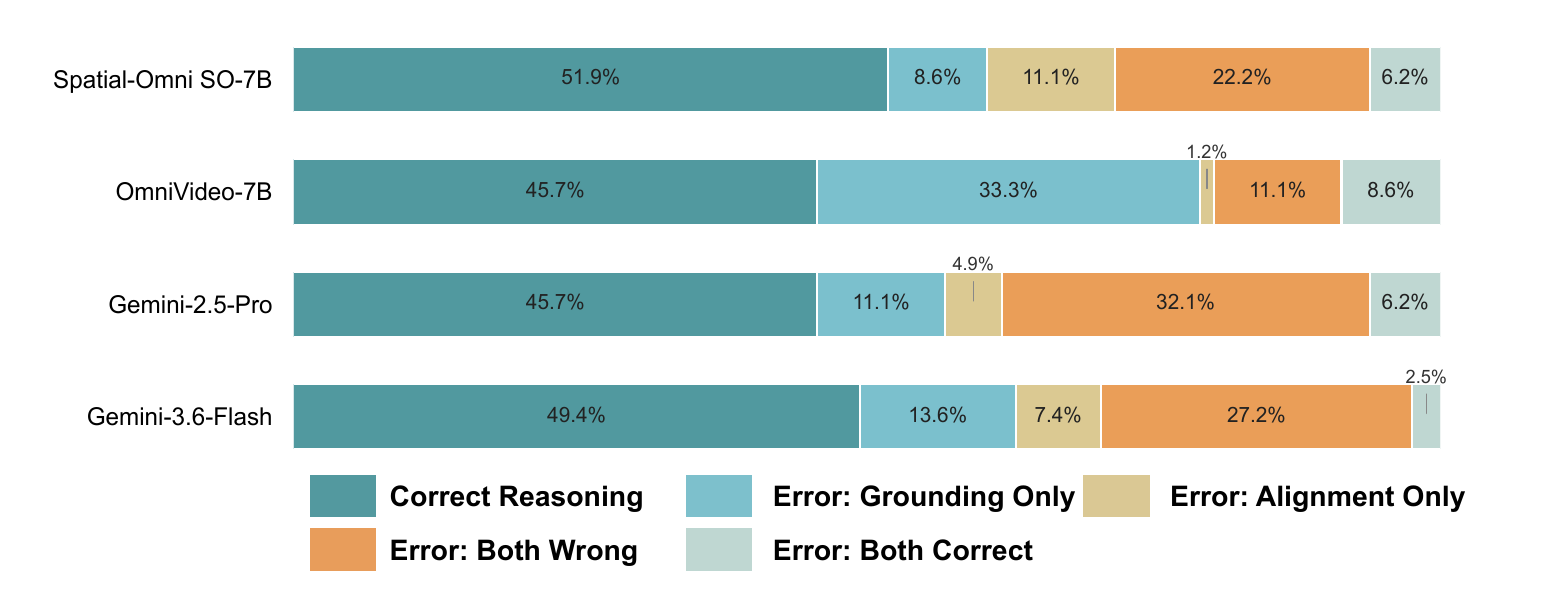}
        \caption{Quantitative diagnostic results across representative models.}
        \label{fig:savu_diag_quant}
    \end{subfigure}

    \vspace{0.7em}

    \begin{subfigure}[t]{0.85\textwidth}
        \centering
        \includegraphics[width=\linewidth]
        {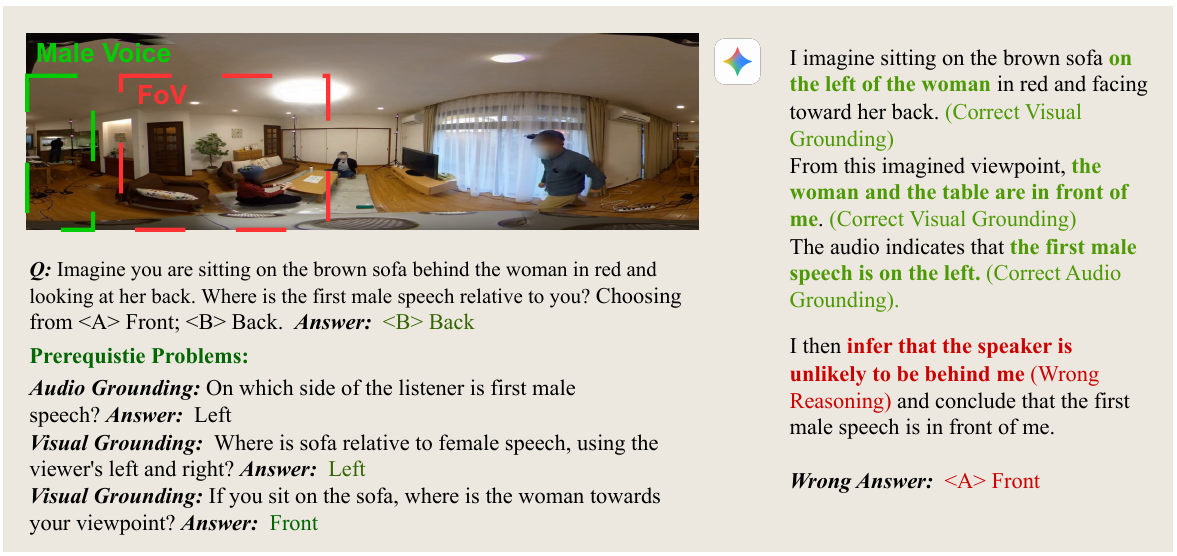}
        \caption{A representative qualitative case.}
        \label{fig:savu_diag_qual}
    \end{subfigure}

    \caption{
    The analysis of SAVU-Diag: (a) distribution of reasoning outcomes by grounding and alignment correctness across representative models and (b) a representative reasoning failure with prerequisite tasks correct
    }
    \label{fig:savu_diag}
\end{figure*}

\paragraph{Quantitative Analysis.}
SAVU-Diag reveals distinct bottlenecks behind similar reasoning accuracy. As shown in Fig.~\ref{fig:savu_diag_quant}, OmniVideo-7B is strongly grounding-limited: grounding-only failures account for 61.4\% of its incorrect reasoning cases. Spatial-Omni shows a different pattern. Only 17.9\% of its reasoning errors are grounding-only, while 23.1\% involve alignment-only failures and 46.2\% involve failures in both prerequisites. This shift suggests that, for a model with stronger spatial-audio specialization, the dominant bottleneck moves beyond basic grounding toward establishing reliable cross-modal relations. The Gemini models exhibit another pattern, with most reasoning errors accompanied by simultaneous grounding and alignment failures. Moreover, these models still produce reasoning errors when both evaluated prerequisites are answered correctly, indicating that successful prerequisite acquisition does not guarantee successful relation composition. Overall, SAVU-Diag shows that similar reasoning accuracy can conceal substantially different bottlenecks in grounding, cross-modal alignment, and relation composition.

\paragraph{Qualitative Analysis.}
Fig.~\ref{fig:savu_diag_qual} illustrates a failure case for Gemini-3.6-Flash. In this example, all grounding prerequisites are answered correctly. Nevertheless, the model composes an incorrect spatial relation and reaches the wrong final answer. This case shows that correct prerequisite predictions do not necessarily lead to correct spatial reasoning.

\subsection{Discussion: Can Explicit Spatial Cues Improve Spatial Understanding?}
\begin{figure*}[t]
    \centering

    \begin{subfigure}[t]{0.47\textwidth}
        \centering
        \includegraphics[width=\linewidth]
        {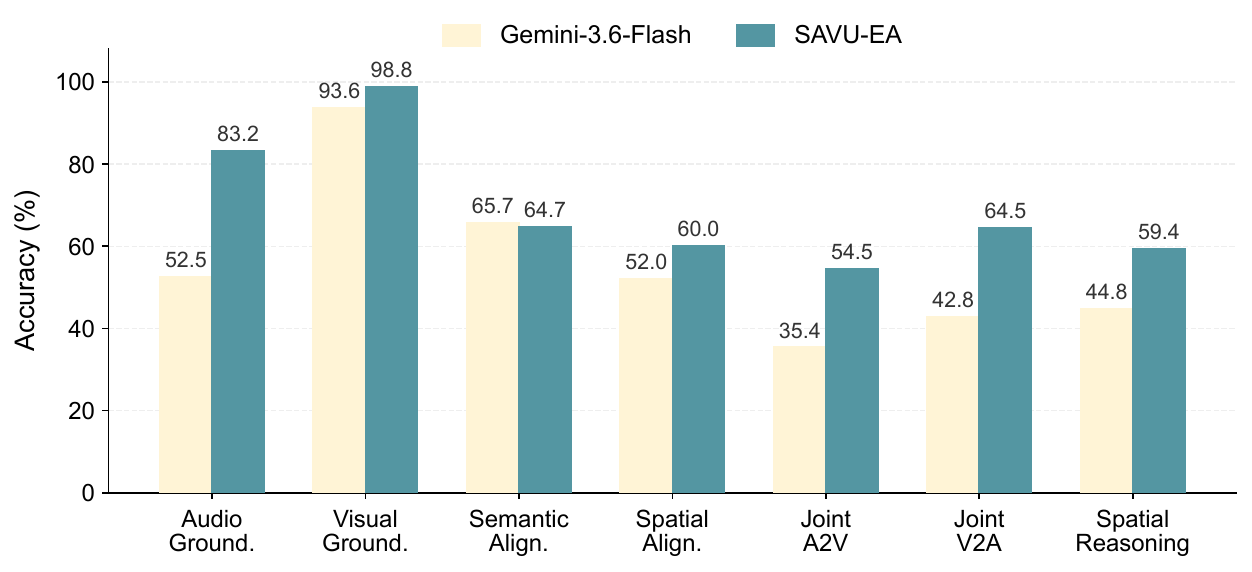}
        \caption{Performance on the full SAVU-Bench.}
        \label{fig:savu_ea_full}
    \end{subfigure}
    \hspace{0.045\textwidth}
    \begin{subfigure}[t]{0.47\textwidth}
        \centering
        \includegraphics[width=\linewidth]
        {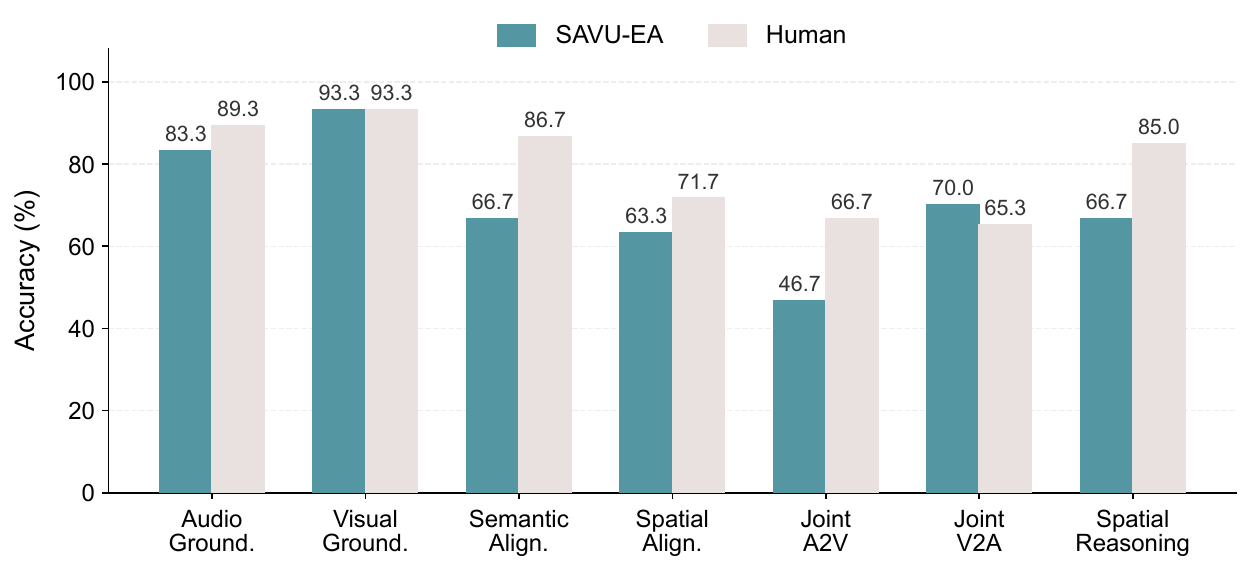}
        \caption{Comparison with humans on the human subset.}
        \label{fig:savu_ea_human}
    \end{subfigure}

    \caption{
        The results of SAVU-EA: (a) comparison with Gemini-3.6-Flash on the full SAVU-Bench and (b) comparison with human performance on the human-evaluation subset.
    }
    \label{fig:savu_ea_results}
\end{figure*}

The previous results reveal two complementary limitations. SAVU-Bench shows that auditory spatial grounding remains weak, while SAVU-Diag indicates that reasoning errors can arise both with and without failures in prerequisite tasks. This motivates a natural question: how much can current models benefit from more explicit spatial cues?

To investigate this, we introduce SAVU-EA, a simple training-free baseline that augments the model input with predicted spatial information. OpenVocabularySELD extracts sound events and their spatial trajectories from FOA audio, while Qwen3.5-Omni-Plus extracts visible objects, positions, and motion from silent video. These predicted cues are then provided to Gemini-3.6-Flash together with the original inputs and question. For perception tasks, SAVU-EA uses only the cues and inputs from the corresponding modality.

As shown in Fig.~\ref{fig:savu_ea_full}, SAVU-EA substantially improves several spatial tasks on the full benchmark. Compared with the original Gemini-3.6-Flash, Audio Grounding improves by 30.7 percentage points, while Joint A2V and Joint V2A increase by 19.1 and 21.7 points, respectively. Spatial Reasoning also improves by 14. points. These results show that making spatial information more explicit is particularly helpful when models must localize sound sources or combine spatial and semantic correspondence across modalities.

The human comparison in Fig.~\ref{fig:savu_ea_human} provides a complementary view. SAVU-EA approaches or exceeds human performance on Visual Grounding and Joint V2A, but clear gaps remain in Semantic Alignment, Spatial Alignment, Joint A2V, and Spatial Reasoning. In particular, the remaining reasoning gap persists even after spatial grounding is substantially improved.

Together, these results lead to two observations. First, stronger access to spatial cues can substantially improve grounding and cross-modal matching, confirming that perceptual quality is an important bottleneck for current models. Second, better spatial perception does not automatically translate into reliable reasoning. Even when relevant spatial information is made more explicit, models still struggle to preserve and combine semantic and spatial relations into the final spatial judgment.

\section{Conclusion}
We introduced SAVU-Bench, a real-world benchmark for systematically evaluating spatial audio-visual understanding, together with SAVU-Diag for scene-linked diagnosis. Our evaluation reveals a clear capability gap: current models perform well on visual grounding but remain substantially weaker in auditory spatial grounding, joint matching, and spatial reasoning. SAVU-Diag further shows that similar reasoning accuracy can arise from different prerequisite-error patterns, while reasoning failures can persist even when the evaluated grounding and alignment tasks are answered correctly. Finally, SAVU-EA demonstrates that making spatial cues more explicit substantially improves grounding and downstream performance, but does not eliminate the reasoning gap. Overall, these findings suggest that robust spatial audio-visual understanding requires progress in both acquiring reliable spatial relations and consistently combining them for higher-level reasoning.

\section*{AI use statement}

In this work, we used generative AI tools to assist in generating initial candidate questions for our proposed benchmark. All generated candidates were subsequently reviewed and verified by human annotators to ensure their validity. We also used generative AI tools to assist with language polishing in this paper. We take responsibility for the final content of this work, including all text, claims, and artifacts produced with the aid of generative AI.

\subsection*{Ethics statement}

This work involves human annotators and participants for benchmark construction, expert verification, and human evaluation. Participants are recruited only for annotation and evaluation of the audio-visual tasks described in this paper, and the human evaluators are disjoint from those involved in benchmark construction. The benchmark is constructed from existing research datasets, and we will follow the applicable licenses and usage terms of the original data sources when releasing the derived benchmark.

\subsection*{Reproducibility statement}

We provide detailed descriptions of the benchmark construction and evaluation procedures in both the main paper and the appendix. Sec.~\ref{sec:data} describes the overall data curation pipeline, while Appendix~\ref{app:spatial_reference} details the spatial reference construction and FOA transformation. Additional task construction details and representative examples are provided in Appendix~\ref{app:task_details}, and the construction of SAVU-Diag and the human annotation and evaluation protocols are described in Appendix~\ref{app:savu_diag_construction} and ~\ref{app:human_annotation}. We also provide additional benchmark examples in the supplementary material. Upon acceptance, we plan to publicly release the benchmark, evaluation code, and relevant construction scripts to facilitate reproducibility and future research.

\bibliography{iclr2027_conference}
\bibliographystyle{iclr2027_conference}

\newpage
\appendix
\section{Appendix}

\subsection{Experiment Results and Analysis}
\label{sec:complete_results}

\subsubsection{Chance-Level Calculation}
\label{app:chance_calculation}

Under the AA definition in Eq.~\ref{eq:circular_aa}, each question is evaluated under all cyclic permutations of its answer choices, and accuracy is computed over all question--ordering pairs. For Audio Grounding, Visual Grounding, Semantic Alignment, and Spatial Alignment, all questions contain two answer choices, resulting in a chance level of $50.0\%$. Joint A2V and Joint V2A contain three choices, resulting in a chance level of $33.3\%$.

One special case is the spatial reasoning tasks which contains question with different numbers of answer choices. Spatial Reasoning contains 66 two-choice questions and 15 four-choice questions. Its chance level is therefore

\begin{equation}
\mathrm{Chance}
=
\frac{81}{
66 \times 2 + 15 \times 4
}
\times 100\%
=
42.1875\%
\approx 42.2\%.
\end{equation}

\subsubsection{Human Subset}
\label{app:human_subset}

\begin{table}[!htbp]
\centering
\begingroup

\setlength{\tabcolsep}{4.2pt}
\renewcommand{\arraystretch}{1.10}
\footnotesize

\caption{
\textbf{Complete results on the human-evaluation subset of SAVU-Bench.}
Each task contains 30 sampled questions.}
\label{tab:complete_human_results}

\begin{adjustbox}{max width=\linewidth}
\begin{tabular}{lccccccc}
\toprule

\textbf{Model}
& \shortstack{\textbf{Audio}\\\textbf{Grounding}}
& \shortstack{\textbf{Visual}\\\textbf{Grounding}}
& \shortstack{\textbf{Semantic}\\\textbf{Alignment}}
& \shortstack{\textbf{Spatial}\\\textbf{Alignment}}
& \shortstack{\textbf{Joint}\\\textbf{A2V}}
& \shortstack{\textbf{Joint}\\\textbf{V2A}}
& \shortstack{\textbf{Spatial}\\\textbf{Reasoning}}
\\

\midrule

\multicolumn{8}{l}{\textbf{Open-weight Models}} \\

Qwen2.5-Omni-7B
& 66.67
& 76.67
& 56.67
& 53.33
& 36.67
& 26.67
& 46.67 \\

Spatial-Omni SO-7B
& 66.67
& 66.67
& 53.33
& 50.00
& 36.67
& 33.33
& 46.67 \\

MiniCPM-o-4.5
& 56.67
& 86.67
& 53.33
& 53.33
& 26.67
& 20.00
& 43.33 \\

OmniVideo-7B
& 56.67
& 76.67
& 50.00
& 50.00
& 36.67
& 33.33
& 43.33 \\

video-SALMONN2+ 7B
& 50.00
& 86.67
& 53.33
& 50.00
& 40.00
& 43.33
& 33.33 \\

Nemotron-3-Nano-Omni
& 36.67
& 100.00
& 46.67
& 56.67
& 20.00
& 33.33
& 40.00 \\

DOTS-3-Note-Preview
& 56.67
& 100.00
& 50.00
& 56.67
& 30.00
& 33.33
& 30.00 \\

\midrule

\multicolumn{8}{l}{\textbf{Closed-source Models}} \\

Qwen3.5-Omni-Plus
& 66.67
& 100.00
& 73.33
& 46.67
& 33.33
& 20.00
& 40.00 \\

Qwen3.5-Omni-Flash
& 63.33
& 86.67
& 63.33
& 60.00
& 30.00
& 20.00
& 43.33 \\

Gemini-2.5-Pro
& 50.00
& 96.67
& 70.00
& 43.33
& 23.33
& 30.00
& 43.33 \\

Gemini-3.6-Flash
& 56.67
& 96.67
& 70.00
& 43.33
& 36.67
& 43.33
& 33.33 \\

Doubao-Seed-2.0-Lite
& 56.67
& 100.00
& 63.33
& 56.67
& 23.33
& 43.33
& 30.00 \\

\midrule

\multicolumn{8}{l}{\textbf{Evidence-Augmented Model}} \\

SAVU-EA
& 83.33
& 93.33
& 66.67
& 63.33
& 46.67
& 70.00
& 66.67 \\

\midrule

\rowcolor{gray!5}
\textbf{Human}
& \textbf{89.33}
& \textbf{93.33}
& \textbf{86.67}
& \textbf{71.67}
& \textbf{66.67}
& \textbf{65.33}
& \textbf{85.00} \\

\bottomrule
\end{tabular}
\end{adjustbox}

\endgroup
\end{table}

\FloatBarrier

The complete human-subset results in Table~\ref{tab:complete_human_results} further distinguish isolated spatial capabilities from cross-modal integration. Visual Grounding is already strong for many models, with several reaching 96.67--100.00\%, comparable to or exceeding the human accuracy of 93.33\%. Some models also perform relatively well on individual alignment tasks. For example, Qwen3.5-Omni-Plus achieves 73.33\% on Semantic Alignment, yet drops to 33.33\%/20.00\% on Joint A2V/V2A. More generally, the best base-model accuracies reach only 40.00\% on Joint A2V, 43.33\% on Joint V2A, and 46.67\% on Spatial Reasoning, compared with 66.67\%, 65.33\%, and 85.00\% for humans, respectively. These results show that strong performance on individual perception or alignment tasks does not necessarily translate into reliable integration and reasoning over multiple spatial relations.

SAVU-EA substantially narrows several of these gaps, reaching 83.33\% on Audio Grounding, 70.00\% on Joint V2A, and 66.67\% on Spatial Reasoning. However, clear gaps remain in Semantic Alignment, Joint A2V, and especially Spatial Reasoning relative to human performance. This comparison further suggests that making spatial evidence more explicit can improve downstream performance, while reliable reasoning still requires models to consistently integrate and compose cross-modal spatial relations.

\subsubsection{SAVU-Diag}
\label{app:diag_results}

Table~\ref{tab:savu_diag_complete} further reveals that similar reasoning accuracy can arise from substantially different prerequisite bottlenecks. For each reasoning instance, \(G_1\) indicates that all associated Grounding prerequisite questions are answered correctly, whereas \(G_0\) indicates at least one incorrect Grounding answer. Similarly, \(A_1\) requires all associated Semantic Alignment and Spatial Alignment questions to be answered correctly, whereas \(A_0\) indicates at least one incorrect Alignment answer.

Spatial-Omni and Nemotron-3-Nano-Omni, for example, achieve the same reasoning accuracy of 51.85\%, yet exhibit markedly different error profiles. Among their incorrect reasoning cases, Spatial-Omni has relatively few grounding-only failures (17.9\%) but substantially more alignment-only (23.1\%) and joint prerequisite failures (46.2\%). In contrast, Nemotron is more strongly associated with grounding-only failures (41.0\%), while joint prerequisite failures account for only 28.2\%. A similar contrast appears between MiniCPM-o-4.5 and Qwen3.5-Omni-Plus, which both achieve 48.15\% reasoning accuracy: 50.0\% of MiniCPM's reasoning errors are grounding-only, whereas 52.4\% of Qwen3.5-Omni-Plus errors occur when both grounding and alignment are incorrect.

The complete results also expose more concentrated failure modes. For video-SALMONN2+, 88.9\% of incorrect reasoning cases coincide with failures in both grounding and alignment, suggesting that its reasoning errors are largely accompanied by missing prerequisite evidence. At the same time, most models still exhibit reasoning failures when both evaluated prerequisites are answered correctly. Together, these results separate two complementary limitations: models differ substantially in how reliably they acquire the prerequisite spatial relations, and correct prerequisite predictions do not necessarily translate into correct relation composition during reasoning.

We also conduct an oracle prerequisite evaluation. We evaluate Gemini-3.6-Flash on all 81 Spatial Reasoning questions under the single-response protocol. In the oracle condition, the original inputs are supplemented with correct grounding and alignment prerequisite facts, without providing the final answer. Accuracy increases from 49.38\% to 77.78\%, a gain of 28.40 percentage points. This result demonstrates the benefit of explicitly providing correct prerequisite information, while the remaining 18 errors show that such information does not ensure a correct final answer in every case.

\begin{table}[!htbp]
\centering
\begingroup

\setlength{\tabcolsep}{5.0pt}
\renewcommand{\arraystretch}{1.10}
\footnotesize

\caption{
\textbf{Complete diagnostic results on SAVU-Diag.}
Results are reported as percentages over all reasoning instances.
$G1/G0$ and $A1/A0$ denote correct/incorrect grounding and
alignment outcomes, respectively.
}
\label{tab:savu_diag_complete}

\begin{adjustbox}{max width=\linewidth}
\begin{tabular}{lccccc}
\toprule

\textbf{Model}
& \textbf{Reasoning Correct}
& \textbf{R Error: G0 A1}
& \textbf{R Error: G1 A0}
& \textbf{R Error: G0 A0}
& \textbf{R Error: G1 A1} \\

\midrule

Qwen2.5-Omni-7B
& 43.21
& 38.27
& 1.23
& 11.11
& 6.17 \\

Spatial-Omni SO-7B
& 51.85
& 8.64
& 11.11
& 22.22
& 6.17 \\

MiniCPM-o-4.5
& 48.15
& 25.93
& 4.94
& 14.81
& 6.17 \\

OmniVideo-7B
& 45.68
& 33.33
& 1.23
& 11.11
& 8.64 \\

video-SALMONN2+ 7B
& 44.44
& 0.00
& 6.17
& 49.38
& 0.00 \\

Nemotron-3-Nano-Omni
& 51.85
& 19.75
& 4.94
& 13.58
& 9.88 \\

DOTS-3-Note-Preview
& 41.98
& 28.40
& 4.94
& 13.58
& 11.11 \\

Qwen-3.5-Omni-Plus
& 48.15
& 7.41
& 8.64
& 27.16
& 8.64
\\

Qwen-3.5-Omni-Flash
& 44.44
& 8.64
& 3.70
& 41.98
& 1.23
\\

Gemini-2.5-Pro
& 45.68
& 11.11
& 4.94
& 32.10
& 6.17 \\

Gemini-3.6-Flash
& 49.38
& 13.58
& 7.41
& 27.16
& 2.47 \\

Doubao-Seed-2.0-Lite
& 46.91
& 13.58
& 12.35
& 22.22
& 4.94 \\

\bottomrule
\end{tabular}
\end{adjustbox}

\endgroup
\end{table}

\FloatBarrier

\subsubsection{Ablation Study: Short-cut Analysis}
\label{app:shortcut_analysis}

\begin{table}[!htbp]
\centering
\begingroup

\setlength{\tabcolsep}{4.8pt}
\renewcommand{\arraystretch}{1.10}
\footnotesize

\caption{
\textbf{Modality ablation on Spatial Reasoning.}
Parentheses report the change from Audio+Video in percentage points;
decreases are shown in red.
}
\label{tab:modality_ablation}

\begin{adjustbox}{max width=\linewidth}
\begin{tabular}{llccc}
\toprule

\multirow{2}{*}{\textbf{Type}}
& \multirow{2}{*}{\textbf{Model}}
& \multicolumn{3}{c}{\textbf{Spatial Reasoning}} \\

\cmidrule(lr){3-5}

&
& \textbf{Audio+Video}
& \textbf{Audio-only}
& \textbf{Video-only} \\

\midrule

\multirow{7}{*}{Open-weight}

& Qwen2.5-Omni-7B
& 43.21
& 40.74 {\scriptsize\textcolor{red!70!black}{($-2.47$)}}
& 41.98 {\scriptsize\textcolor{red!70!black}{($-1.23$)}} \\

& Spatial-Omni SO-7B
& 51.85
& 46.91 {\scriptsize\textcolor{red!70!black}{($-4.94$)}}
& 45.68 {\scriptsize\textcolor{red!70!black}{($-6.17$)}} \\

& MiniCPM-o-4.5
& 48.15
& 46.91 {\scriptsize\textcolor{red!70!black}{($-1.23$)}}
& 44.44 {\scriptsize\textcolor{red!70!black}{($-3.70$)}} \\

& OmniVideo-7B
& 45.68
& 44.44 {\scriptsize\textcolor{red!70!black}{($-1.23$)}}
& 44.44 {\scriptsize\textcolor{red!70!black}{($-1.23$)}} \\

& video-SALMONN2+ 7B
& 44.44
& 35.80 {\scriptsize\textcolor{red!70!black}{($-8.64$)}}
& 41.98 {\scriptsize\textcolor{red!70!black}{($-2.47$)}} \\

& Nemotron-3-Nano-Omni
& 51.85
& 45.68 {\scriptsize\textcolor{red!70!black}{($-6.17$)}}
& 48.15 {\scriptsize\textcolor{red!70!black}{($-3.70$)}} \\

& DOTS-3-Note-Preview
& 41.98
& 40.74 {\scriptsize\textcolor{red!70!black}{($-1.23$)}}
& 38.27 {\scriptsize\textcolor{red!70!black}{($-3.70$)}} \\

\midrule

\multirow{5}{*}{Closed-source}

& Qwen3.5-Omni-Plus
& 48.15
& 45.68 {\scriptsize\textcolor{red!70!black}{($-2.47$)}}
& 44.44 {\scriptsize\textcolor{red!70!black}{($-3.70$)}} \\

& Qwen3.5-Omni-Flash
& 44.44
& 43.21 {\scriptsize\textcolor{red!70!black}{($-1.23$)}}
& 41.98 {\scriptsize\textcolor{red!70!black}{($-2.47$)}} \\

& Gemini-3.6-Flash
& 49.38
& 40.74 {\scriptsize\textcolor{red!70!black}{($-8.64$)}}
& 45.68 {\scriptsize\textcolor{red!70!black}{($-3.70$)}} \\

& Gemini-2.5-Pro
& 45.68
& 44.44 {\scriptsize\textcolor{red!70!black}{($-1.23$)}}
& 39.51 {\scriptsize\textcolor{red!70!black}{($-6.17$)}} \\

& Doubao-Seed-2.0-Lite
& 46.91
& 34.57 {\scriptsize\textcolor{red!70!black}{($-12.35$)}}
& 43.21 {\scriptsize\textcolor{red!70!black}{($-3.70$)}} \\

\bottomrule
\end{tabular}
\end{adjustbox}

\endgroup
\end{table}

\FloatBarrier

Table~\ref{tab:modality_ablation} compares full audio-visual input with audio-only and video-only inputs on the same Spatial Reasoning questions. All 12 models achieve higher accuracy with both modalities than with either modality alone. Removing vision causes the largest decreases for Doubao-Seed-2.0-Lite (12.35 percentage points), Gemini-3.6-Flash (8.64 points), and video-SALMONN2+ (8.64 points). Removing audio has the largest effect on Spatial-Omni and Gemini-2.5-Pro, both decreasing by 6.17 points. Averaged across models, removing vision and audio reduces accuracy by 4.32 and 3.50 points, respectively. These results support complementary contributions from audio and vision to Spatial Reasoning, while showing that the relative contribution of each modality varies across models.

\subsection{Detailed Benchmark Construction}
\label{app:benchmark_details}

\subsubsection{Spatial Reference Construction}
\label{app:spatial_reference}

Spatial questions require a consistent reference frame between the visual input and spatial audio. For datasets with panoramic $360^\circ$ video, we first construct a limited-FoV candidate view around the target direction and, when required, align the corresponding audio reference frame with the selected view. For MRSAudio, which already provides limited-FoV videos and their associated spatial audio, we directly use the provided views without additional panoramic rendering.

\paragraph{Limited-FoV visual views.}
For panoramic recordings, we first obtain a target azimuth for each candidate sound event. When source-level spatial annotations are available, we use the provided direction; otherwise, we use the azimuth predicted by OpenVocabularySELD as a pseudo-label. This direction serves as the initial center for constructing a limited-FoV candidate view.

Given the target direction, we render a rectilinear view with a horizontal field of view of $90^\circ$ at a resolution of $960\times540$. The view center is specified by yaw and pitch, with roll fixed to zero. Given the $16{:}9$ aspect ratio, the corresponding vertical field of view is approximately $58.72^\circ$. The source projection is determined from the spherical-video metadata or an explicit source declaration, distinguishing equirectangular panoramas from equi-angular cubemaps.

This procedure produces an initial candidate view centered approximately on the target region rather than a final benchmark annotation. The candidate views are subsequently inspected and refined during human annotation, where the viewing direction, target visibility, and supporting spatial relations are verified. We describe this refinement procedure in Sec.~\ref{app:human_annotation}. Limiting the visual FoV affects only the visual input; the corresponding FOA recording retains the complete acoustic scene, including sound sources outside the rendered view.

\paragraph{FOA convention and horizontal recentering.}
Benchmark spatial audio is represented using FOA,
\[
\mathbf{a}(t)
=
[W(t),Y(t),Z(t),X(t)]^{\mathsf T}.
\]
The positive $X$, $Y$, and $Z$ axes point forward, left, and upward, respectively. Azimuth is measured from the frontal direction and increases toward the left.

For each instance, the audio either retains its original microphone-centered reference frame or is horizontally recentered to the selected visual view. Let $\theta$ denote the recentering yaw. For a view-centered input, the transformed FOA signal is
\[
\mathbf{a}'(t)
=
\mathbf{R}_{\mathrm{ref}}(\theta)\mathbf{a}(t),
\]
where
\[
\mathbf{R}_{\mathrm{ref}}(\theta)
=
\begin{bmatrix}
1 & 0 & 0 & 0\\
0 & \cos\theta & 0 & -\sin\theta\\
0 & 0 & 1 & 0\\
0 & \sin\theta & 0 & \cos\theta
\end{bmatrix}.
\]

This transformation changes the horizontal reference direction while preserving the recorded sound field. Under this convention, a source originally located at azimuth $\phi$ is represented after recentering at
\[
\phi'
=
\operatorname{wrap}(\phi-\theta),
\]
where $\operatorname{wrap}(\cdot)$ maps angles to $(-180^\circ,180^\circ]$.

For view-centered inputs, $\theta$ is set to the selected visual yaw, producing zero relative yaw between the visual center and the transformed acoustic reference. A source located at the selected viewing direction therefore becomes frontal after transformation.

This operation changes only the acoustic reference direction and does not crop the sound field. Restricting the visual input to a limited FoV removes visual access to off-screen regions, while all recorded sound sources remain present in the FOA signal. This property is particularly important for reasoning instances that relate visible entities to out-of-view sound sources. The transformation does not change the listener position.

\subsubsection{Detailed Task Construction}
\label{app:task_details}

This section provides additional details on the task construction described in Sec.~\ref{sec:taxonomy}, including input settings, negative construction, and the spatial relations required by the reasoning tasks. We will provide data samples for each task in our supplementary materials.

\paragraph{Perception.}
Audio Grounding and Visual Grounding are evaluated independently, with each question receiving only the corresponding modality as input. Both task groups contain three question types: Coarse Localization, Relative Position, and Spatial Dynamics. We use human-interpretable spatial categories as the answer space, such as left/right for localization and left-to-right/right-to-left for motion. This design enables a fair comparison between human and model performance. Humans can reliably recognize coarse spatial relations from natural audio-visual scenes, but are generally not expected to estimate an exact azimuth in degrees. For example, a listener may readily identify a speaker as being on the left, while directly estimating the source at a specific angle such as $30^\circ$ is considerably less natural. We therefore formulate the benchmark around coarse spatial judgments that can be answered consistently by both humans and models.

\begin{figure}[!htbp]
    \centering

    \begin{subfigure}[t]{0.32\linewidth}
        \centering
        \includegraphics[width=\linewidth]
        {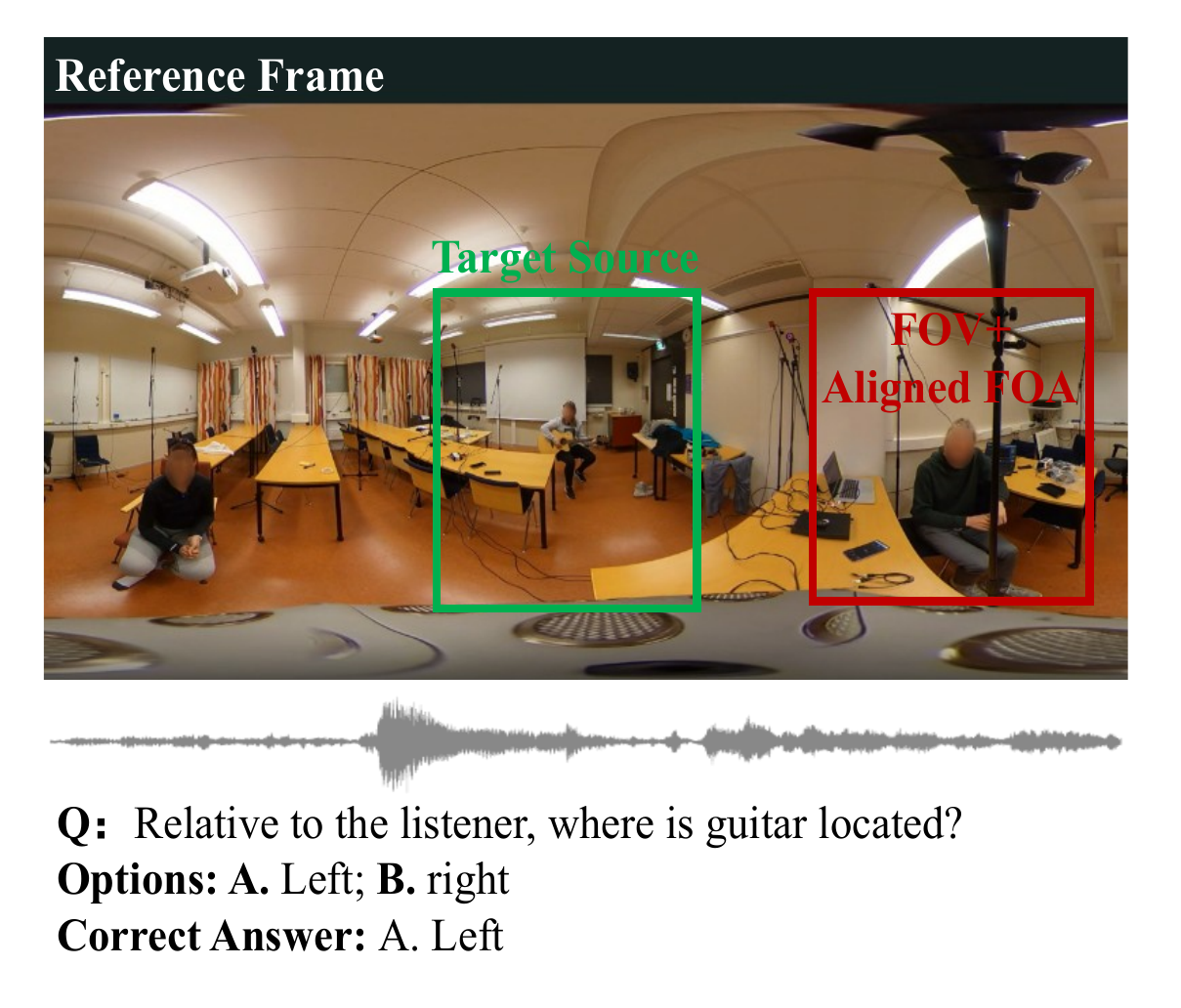}
        \caption{Coarse Localization.}
        \label{fig:coarse_audio}
    \end{subfigure}
    \hfill
    \begin{subfigure}[t]{0.32\linewidth}
        \centering
        \includegraphics[width=\linewidth]
        {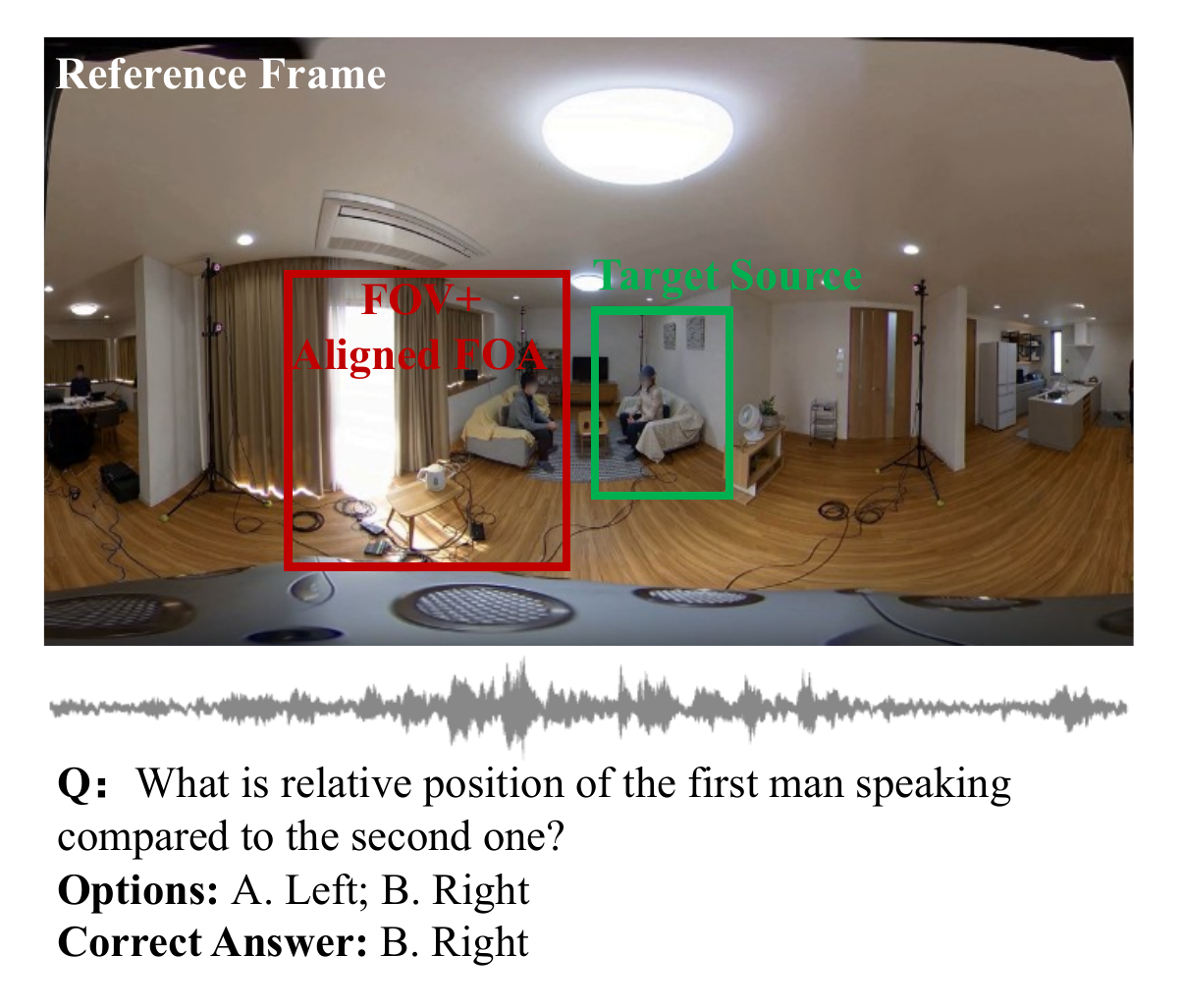}
        \caption{Relative Position.}
        \label{fig:relative_audio}
    \end{subfigure}
    \hfill
    \begin{subfigure}[t]{0.32\linewidth}
        \centering
        \includegraphics[width=\linewidth]
        {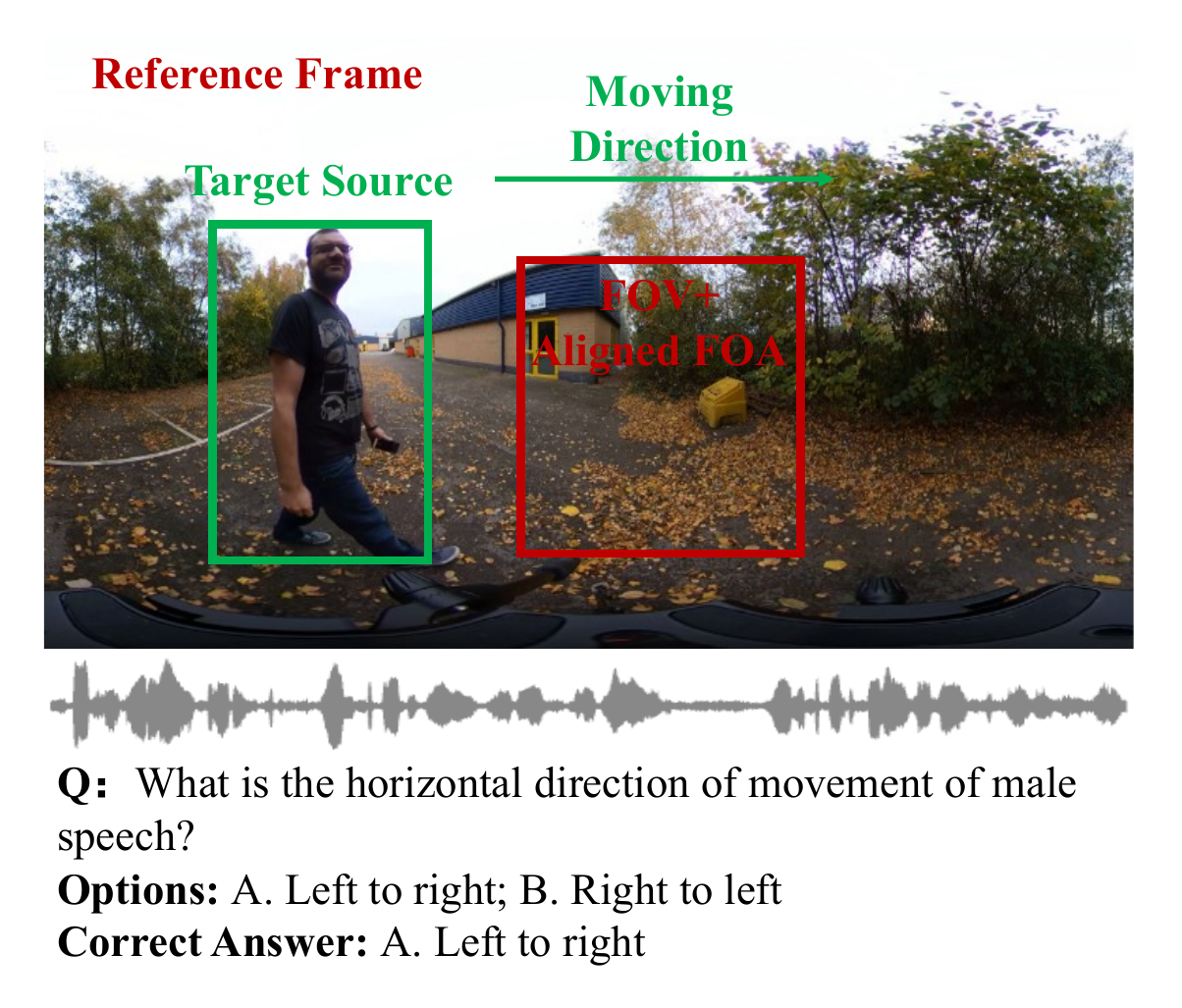}
        \caption{Spatial Dynamics.}
        \label{fig:dynamic_audio}
    \end{subfigure}

    \caption{
    Representative examples of the three Audio Grounding question types:
    (a) Coarse Localization, (b) Relative Position, and
    (c) Spatial Dynamics.
    }
    \label{fig:audio_perception_examples}
\end{figure}

For Audio Grounding, Coarse Localization asks for the coarse direction of a target sound source, such as whether it is located on the left or right. Relative Position further evaluates the spatial relation between two sound sources within the same acoustic scene. To avoid ambiguous comparisons, the selected sources are required to occupy clearly distinguishable coarse directions rather than the same directional region. Each candidate is subsequently verified manually to ensure that the intended relation is unambiguous. Spatial Dynamics evaluates how the position of a target sound source changes over time. We retain only trajectories with sufficiently clear directional changes and exclude cases involving small or ambiguous movements. Representative examples of the three question types are shown in Fig.~\ref{fig:audio_perception_examples}. 

Visual Grounding follows the same structure as Audio Grounding. Coarse Localization asks for the coarse location of a target object within the selected visual field, such as whether it appears on the left or right. Relative Position evaluates the spatial relation between the target and another visible entity in the same scene. To reduce ambiguity, we retain only pairs whose relative positions are visually distinguishable and remove cases in which the two entities occupy nearly the same spatial region.

\begin{figure}[!htbp]
    \centering
    \includegraphics[width=0.65\linewidth]
    {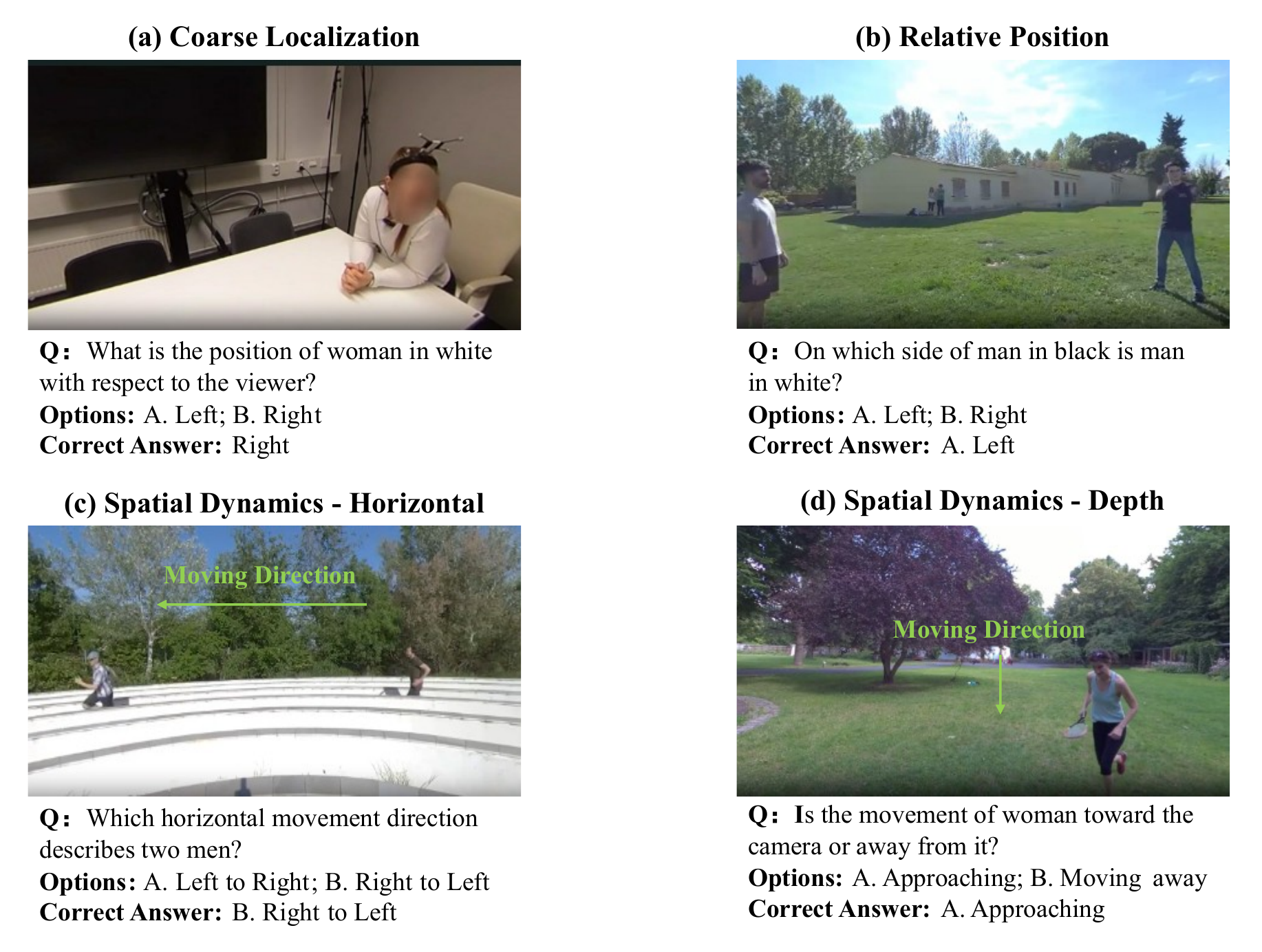}
    \caption{Representative examples of Visual Grounding Types: (a) Coarse Localization, (b) Relative Position, and (c) Spatial Dynamics-Horizontal, (d) Spatial Dynamics-Depth
    }
    \label{fig:visual_perception_examples}
\end{figure}

Spatial Dynamics evaluates how the visual position of a target changes over time. We consider both horizontal motion and motion along the depth direction. For horizontal motion, questions describe clear left-to-right or right-to-left trajectories. For depth motion, the target is required to exhibit an observable change toward or away from the camera. Small or ambiguous movements are excluded during annotation and manual verification. Representative examples of the three question types are shown in Fig.~\ref{fig:visual_perception_examples}. 

Across both Audio Grounding and Visual Grounding, the queried entity is identified through semantic cues, such as a sound category, object category, or visual attribute. The model must therefore first identify the intended target and then recover its spatial state. This design ensures that the tasks evaluate spatial grounding of a semantically specified entity rather than spatial recognition in isolation.

\paragraph{Semantic Alignment.}
Semantic Alignment evaluates whether an audible event is semantically compatible with the visible sound-producing event. Positive pairs contain semantically matched auditory and visual events. 
Negative pairs are constructed by replacing one modality with a segment containing a different sound event, thereby breaking semantic correspondence while preserving the same evaluation format. This construction focuses the task on whether the model can establish the correct event-level correspondence across audio and vision.

\paragraph{Spatial Alignment.}
Spatial Alignment evaluates whether the direction or trajectory of a sound source is consistent with the position or motion of its visual counterpart. To isolate spatial correspondence from semantic mismatch, spatial negatives preserve category-level semantic compatibility while changing the underlying spatial relation.

\begin{figure}[!htbp]
    \centering
    \includegraphics[width=0.65\linewidth]
    {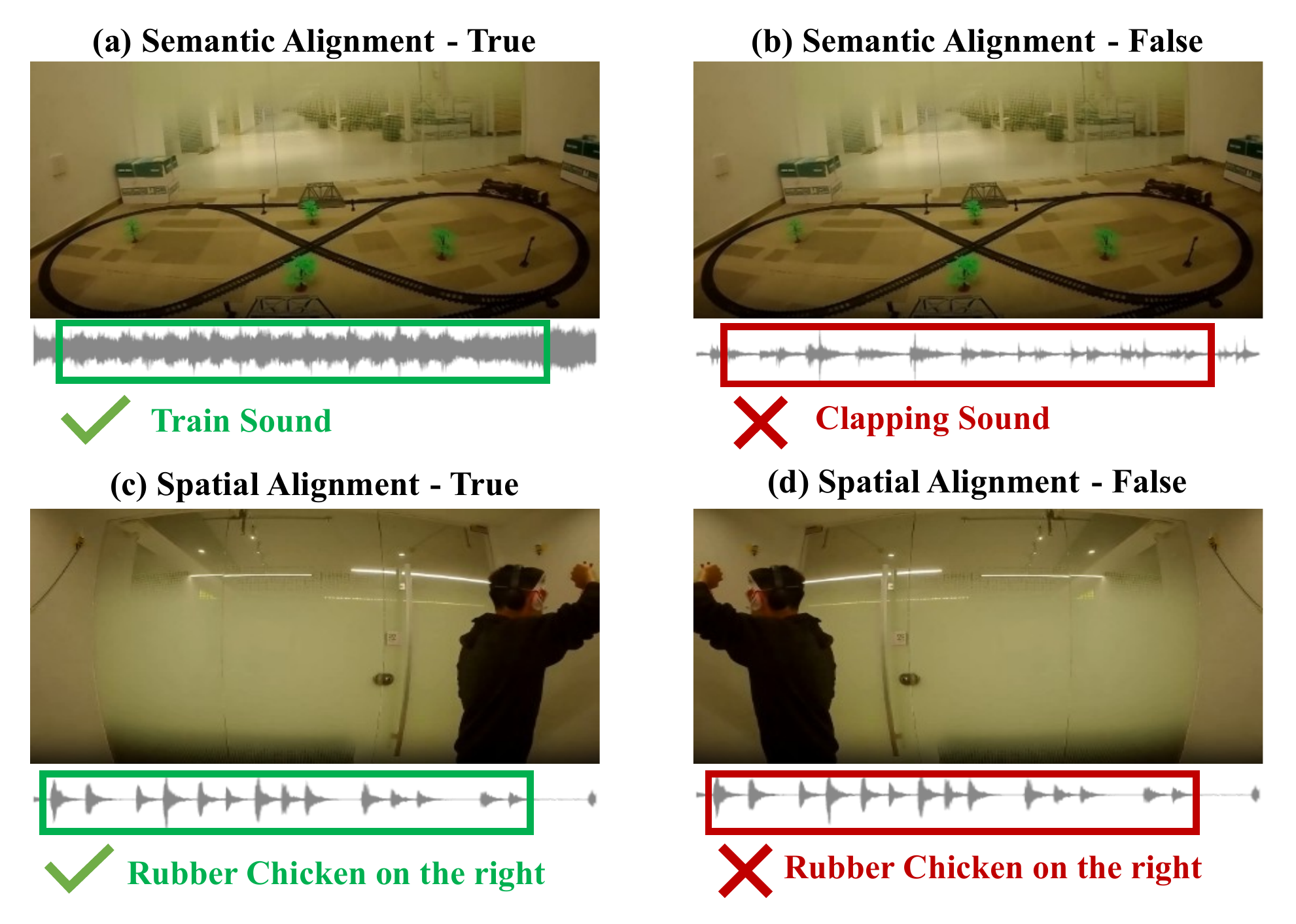}
    \caption{Representative examples of Alignment.
    }
    \label{fig:alignment_examples}
\end{figure}

For FOA recordings, spatially inconsistent pairs can be constructed by rotating the sound field in azimuth while keeping the underlying acoustic event unchanged. We also use segments from the same recording that contain the same target category but exhibit a different spatial position or trajectory. 
In both cases, the semantic identity of the target is preserved while its spatial correspondence with the visual input is altered. This design allows Spatial Alignment to evaluate spatial consistency without conflating it with semantic mismatch. Several representative examples are shown in Fig.~\ref{fig:alignment_examples}.

\paragraph{Joint Matching.}
Joint Matching evaluates whether semantic and spatial correspondence can be satisfied simultaneously. We instantiate the task in both Audio-to-Video (A2V) and Video-to-Audio (V2A) directions.

For A2V, an audio query is paired with three candidate videos: a matched candidate, a semantic negative containing an incompatible event, and a spatial negative that preserves the target category but changes its spatial relation. V2A follows the same construction with a video query and three candidate audio segments. Spatial audio negatives are generated using the corresponding acoustic transformation or trajectory variation described at \ref{app:spatial_reference}. The correct candidate must therefore satisfy both semantic and spatial correspondence, while each negative violates one of these two conditions. This design distinguishes models that recognize semantic compatibility from those that can preserve semantic and spatial consistency jointly.

\begin{figure}[!htbp]
    \centering
    \includegraphics[width=0.90\linewidth]
    {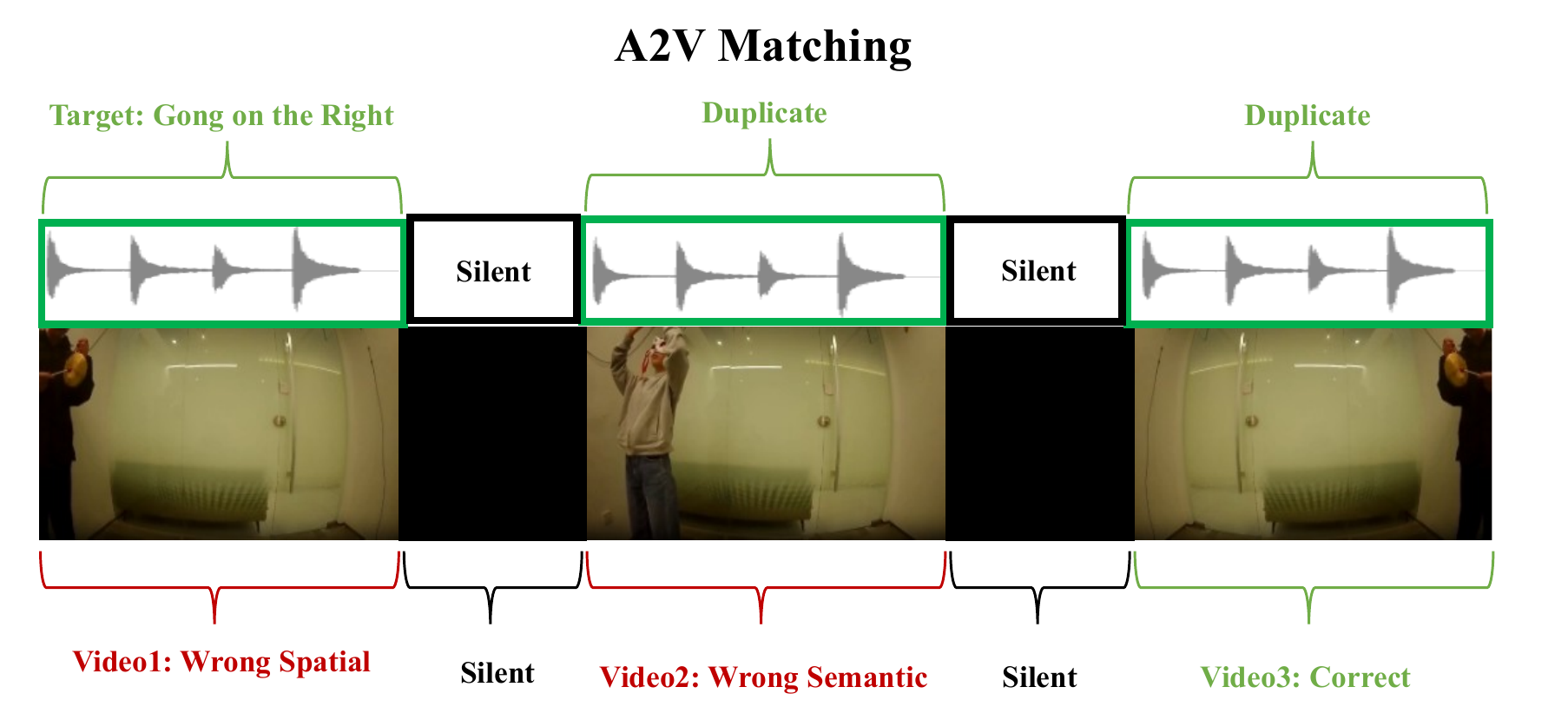}
    \caption{A representative example of Audio-to-Visual (A2V) Matching Task.
    }
    \label{fig:joint_matching}
\end{figure}

Because not all evaluated models support multiple audio or video inputs within a single query, we convert the three candidates into one unified audio-visual sequence for consistent evaluation. For V2A, the query video is repeated three times, with each repetition paired with one audio candidate. Consecutive candidate pairs are separated by a 2-second silent black-screen interval. The three audio candidates are concatenated in the same order, with corresponding 2-second silent intervals between them. A2V follows the same construction with the modalities reversed. This produces a single synchronized audio-video input containing all three candidate pairs, allowing the same evaluation format to be used across models with different input interfaces. A representative sample is shown in Fig.~\ref{fig:joint_matching}.

\paragraph{Spatial Reasoning.}
Spatial Reasoning evaluates whether a model can combine spatial relations across audio and vision to derive a new relation. We consider Relative Direction, Spatial Ordering, and Allocentric Reasoning.

A central design principle of our reasoning tasks is to distribute the required evidence across audio and vision. Each reasoning instance is constructed around three types of entities: a silent visual anchor, an in-view sounding source, and one or more out-of-view sound sources. Among them, the silent anchor plays a key role in linking the two modalities.

Specifically, each question starts from a silent object visible in the scene, together with an active sound-producing source within the FoV. The visual input provides the semantic identities of these entities and their spatial relation. Meanwhile, one or more target sound sources remain outside the FoV and can therefore only be localized through spatial audio. The in-view sounding source serves as the common reference between the two modalities: vision establishes its relation to the silent anchor, while spatial audio establishes its relation to the out-of-view target source.

A question asking for the relation between the silent anchor and an out-of-view sound source therefore requires these two relations to be connected. The silent anchor is accessible only through vision, the out-of-view source is accessible only through audio, and the in-view sounding source bridges the visual and acoustic spatial layouts. This construction reduces direct single-modality shortcuts and encourages the model to combine semantic and spatial evidence from both modalities when deriving the final answer.

\begin{figure}[!htbp]
    \centering
    \includegraphics[width=0.70\linewidth]
    {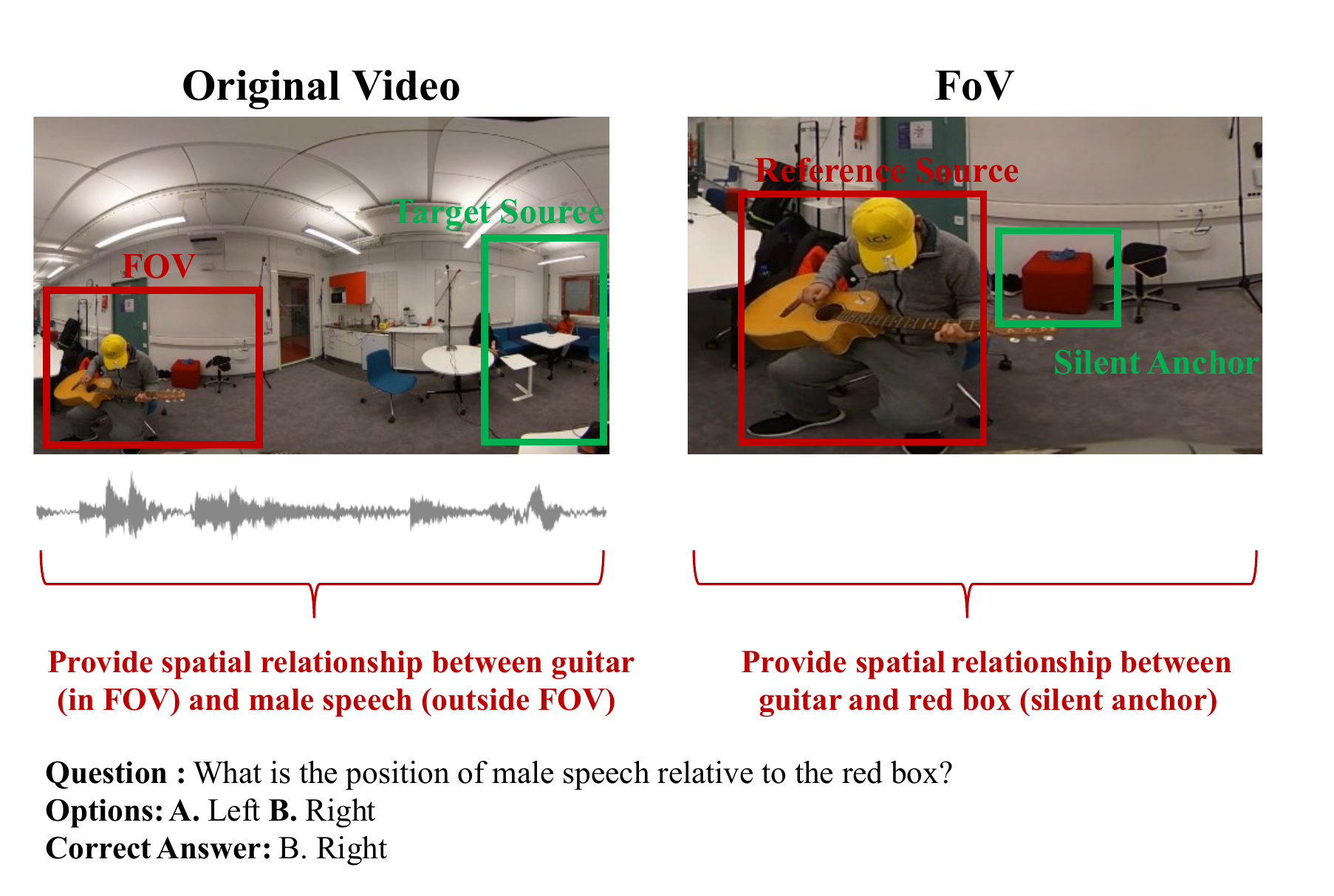}
    \caption{Representative examples of Relative Direction for Spatial Reasoning Task.
    }
    \label{fig:reasoning1}
\end{figure}

\paragraph{Relative Direction.}
Relative Direction evaluates whether a model can compose a pairwise spatial relation across the two modalities. The question asks for the position of an out-of-view sound source relative to a visual reference. For example, as shown in Fig~\ref{fig:reasoning1}, we consider a scene containing a silent red box, an in-view guitarist, and an out-of-view male speaker. Vision provides the spatial relation between the red box and the guitarist, while spatial audio provides the relation between the guitar sound and the male speech. Using the guitarist as the common reference, the model must combine these two relations to infer the position of the male speaker relative to the red box.

In addition to pairwise spatial relations, these instances use verified FoV membership as a spatial constraint. In the illustrated example, the target source lies to the right of the visible guitarist and outside the horizontal FoV. Since both the guitarist and the red box lie inside the FoV, the target must lie beyond its right boundary, placing it to the right of the red box.

\paragraph{Spatial Ordering.}
Spatial Ordering extends pairwise relation composition to multiple spatially distributed sound sources. Starting from a visual anchor, the model is asked to determine the order in which the sources would be encountered when rotating in place from the original observation position. The visual input establishes the relation between the anchor and an in-view sounding source, while spatial audio provides the relative directions of the remaining sound sources, including those outside the FoV. The in-view source therefore serves as the common reference between the visual and acoustic layouts, and the model must combine multiple spatial relations to recover a consistent ordering of the sources.

\begin{figure}[!htbp]
    \centering
    \includegraphics[width=0.70\linewidth]
    {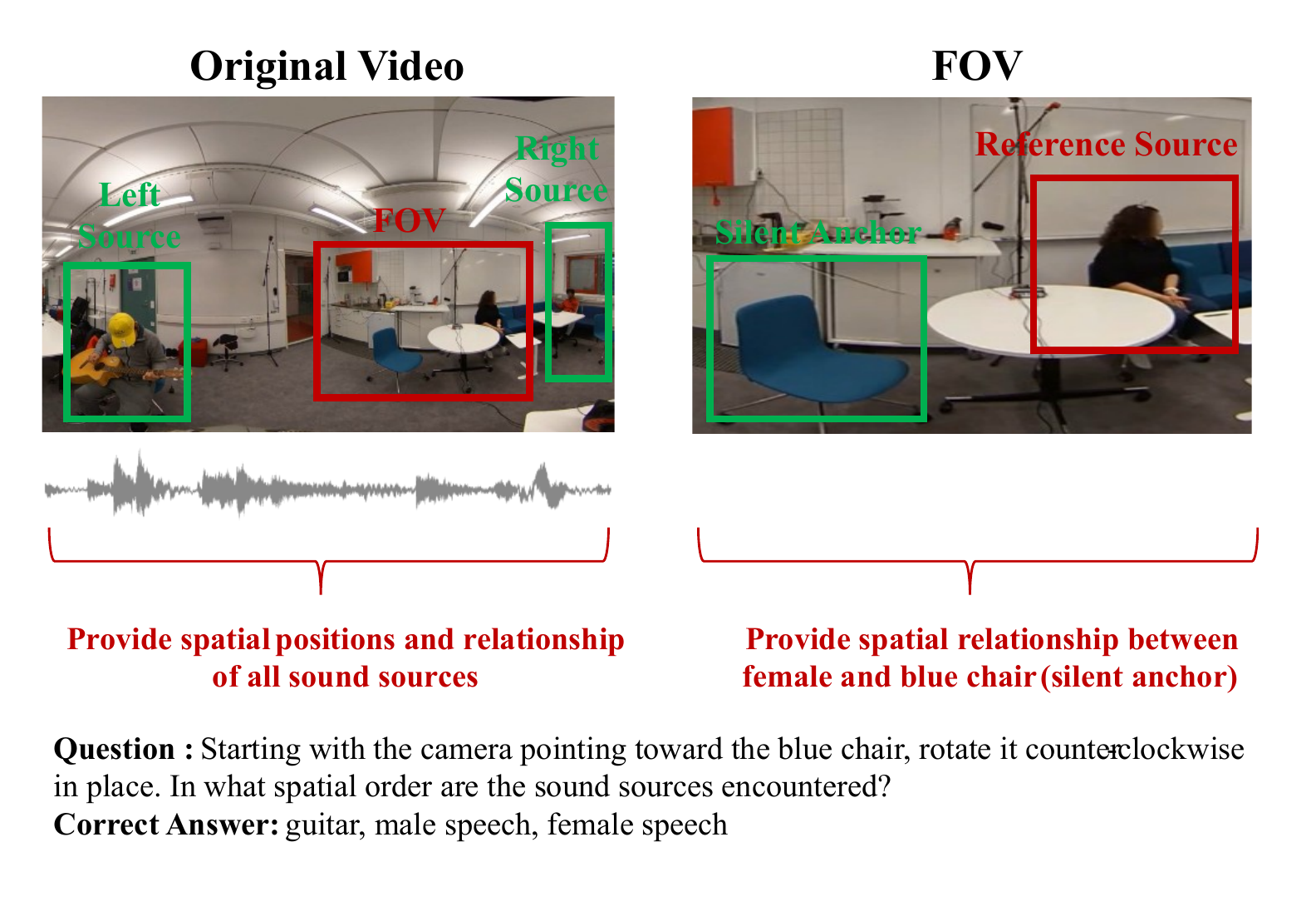}
    \caption{Representative examples of Spatial Ordering for Spatial Reasoning Task.
    }
    \label{fig:reasoning2}
\end{figure}

Because these questions may involve several sound sources, we explicitly control source separability during data construction. Most instances contain at least three sounding sources, including an in-view source and sources located on different sides of the acoustic scene. We preferentially select sources from different event categories so that their semantic identities remain distinguishable. This avoids ambiguous cases in which multiple spatially separated sources share the same event label and therefore become indistinguishable in the answer choices. For instances with additional sources on the same coarse side, we retain only cases with sufficiently separated spatial positions. Candidate relations are first checked using their azimuth information and are subsequently verified during human annotation to remove samples with ambiguous source ordering. A representative example is shown in Fig.~\ref{fig:reasoning2}.

\paragraph{Allocentric Reasoning.}
Allocentric Reasoning further requires the model to reinterpret the scene under a new spatial reference frame. Each question specifies a virtual observer position and a new facing direction using identifiable entities in the scene. The model must first recover the relevant semantic and spatial relations in the original audio-visual layout and then transform these relations into the specified virtual viewpoint.

\begin{figure}[!htbp]
    \centering
    \includegraphics[width=0.70\linewidth]
    {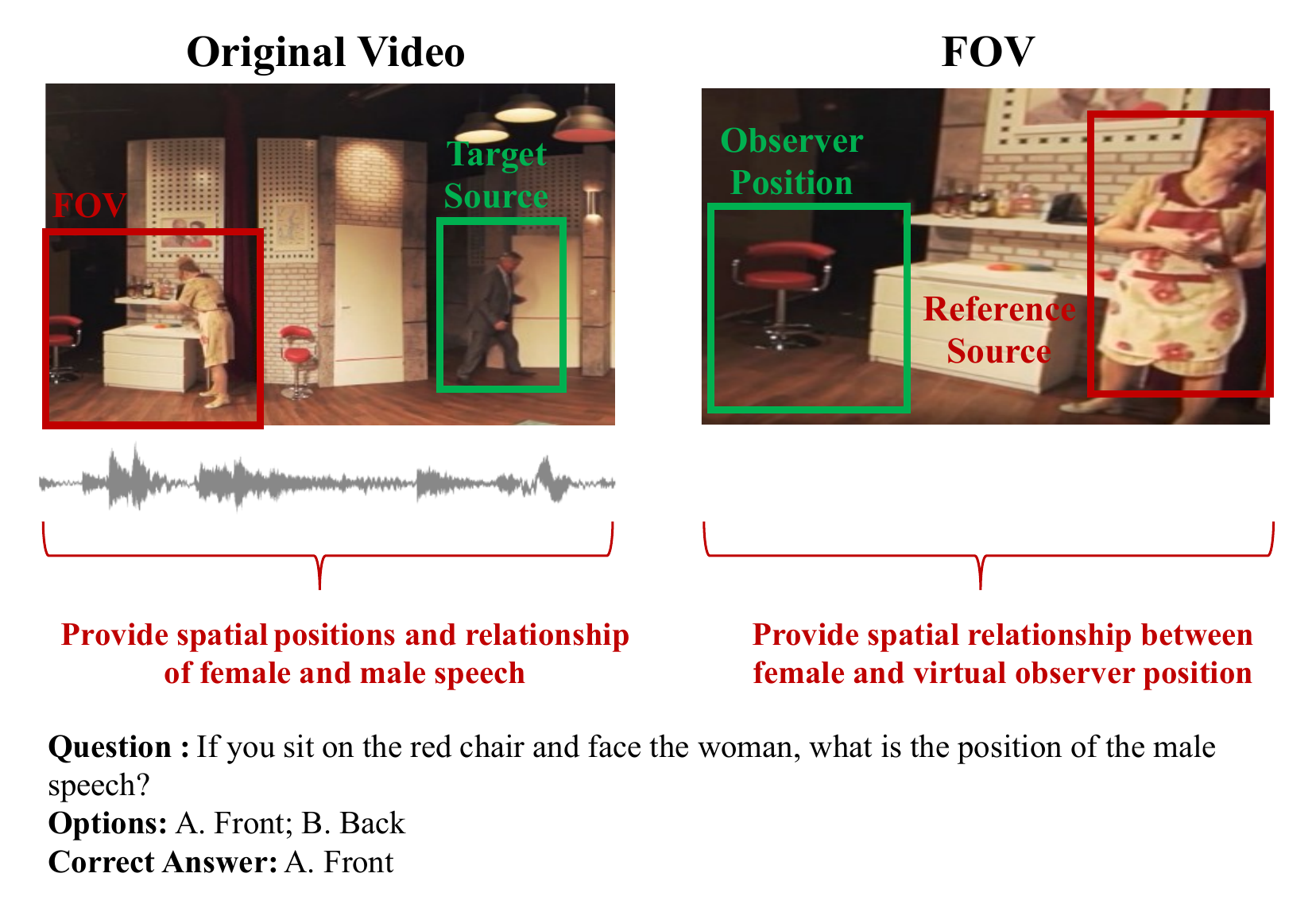}
    \caption{Representative examples of Allocentric Reasoning for Spatial Reasoning Task.
    }
    \label{fig:reasoning3}
\end{figure}

The new reference frame is constructed so that the required visual and acoustic evidence remains connected across modalities. For example, as illustrated in Fig.~\ref{fig:reasoning3}, a question may place the virtual observer at a visible reference and ask the observer to face an in-view sounding source, such as a woman speaking. The visual input identifies the reference entities and establishes the facing direction, while spatial audio provides the relation between the in-view source and another target sound source, such as an out-of-view male speaker. The model must combine these relations and reinterpret the target location from the new observer frame. This construction ensures that the virtual viewpoint is grounded in observable scene entities and that the answer can be derived from complementary visual and spatial-audio evidence.

\subsubsection{Detailed SAVU-Diag Construction}
\label{app:savu_diag_construction}

SAVU-Diag is derived from verified Spatial Reasoning instances. For each reasoning question, we identify the sound sources, visual entities, spatial references, and cross-modal correspondences required to derive the answer. These prerequisite facts and relations are then converted into independent Grounding and Alignment questions.

All questions within a diagnostic group are constructed from the same underlying recording and time interval and preserve the same target identities, visual FoV, and spatial reference-frame conventions. Grounding questions receive only the modality required by the corresponding fact, whereas Alignment and Reasoning questions receive both audio and vision.

The construction starts from the verified reasoning chain associated with each reasoning instance. We provide this chain to GPT-5.4 to identify the grounding facts and cross-modal correspondences required to reach the answer and to generate seed diagnostic questions for these prerequisites. Human annotators then inspect the proposed questions, remove redundant or unsupported prerequisites, and add any missing evidence required by the reasoning process.

Each retained prerequisite is subsequently converted into an independent diagnostic question. Grounding questions evaluate the semantic or spatial facts required within an individual modality, while Alignment questions evaluate the cross-modal correspondences needed to connect these facts. Human annotators independently verify the answer to each diagnostic question and confirm that it is consistent with the original reasoning instance and supported by the corresponding scene evidence.

This construction links each reasoning outcome to the prerequisite capabilities involved in solving the same question, while keeping the diagnostic questions independent from the final reasoning evaluation. It therefore allows us to examine whether a reasoning failure co-occurs with errors in Grounding or Alignment, or persists even when the evaluated prerequisite questions are answered correctly.

\begin{figure}[!htbp]
    \centering
    \includegraphics[width=0.70\linewidth]
    {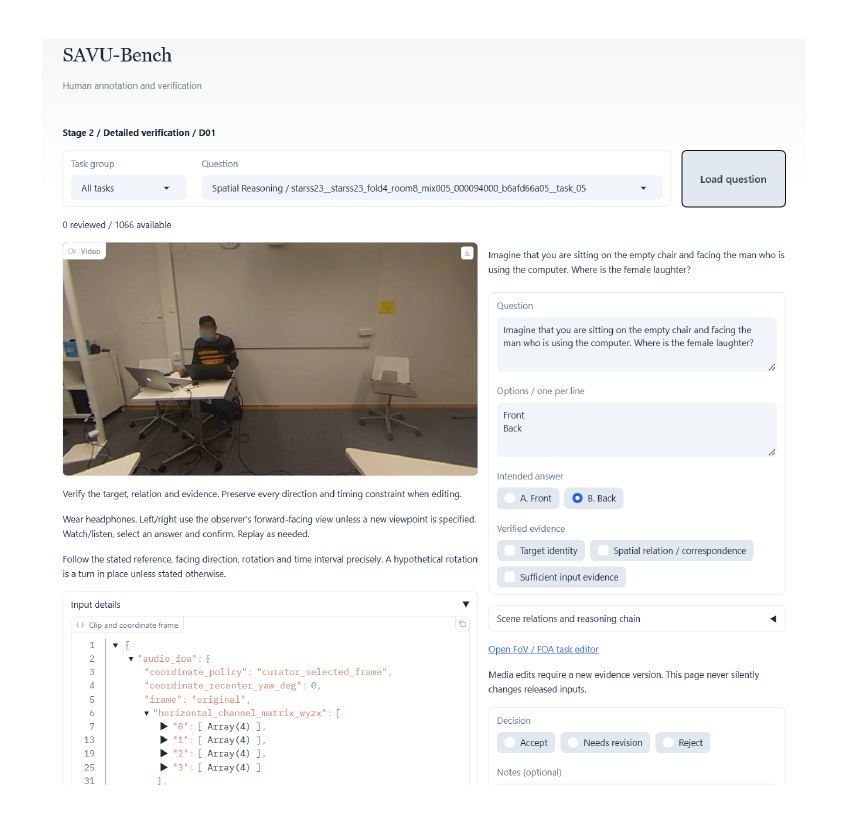}
    \caption{Example of our proposed human annotation platform.
    }
    \label{fig:human_annotation}
\end{figure}

\subsubsection{Human Annotation and Evaluation}
\label{app:human_annotation}

Sec.~\ref{sec:data} summarizes the overall data curation pipeline. Here, we provide additional details on the human annotation procedure, final expert verification, and human evaluation protocol used for SAVU-Bench.

\paragraph{Human Annotation.}
Human annotation is conducted in two stages. In the first stage, ten trained undergraduate annotators perform a coarse verification of the question--answer candidates generated by GPT-5.4 using our annotation platform, as illustrated in Fig.~\ref{fig:human_annotation}. For each candidate, annotators determine whether the question is understandable and solvable from the provided inputs. Only candidates with an acceptance rate above 60\% are retained for the second stage.

The second stage focuses on detailed verification and refinement. We recruit one PhD-level annotator and three undergraduate annotators with experience in spatial-audio-related research. They review the retained question--answer pairs and verify the target identity, spatial relation, and semantic or spatial correspondence required by each question. Candidates with ambiguous targets, unsupported spatial relations, or insufficient evidence are removed. For valid candidates, annotators may refine the visual FoV, FOA orientation, target specification, or question wording so that the intended relation is clearly supported by the input.

Reasoning questions are constructed manually from verified scene relations. For each selected scene, annotators identify the relevant visual anchors, sounding sources, and spatial references and use these relations to define the question and answer. The verified relations are then provided to GPT-5.4 to generate an initial reasoning chain. Human annotators subsequently review and revise the chain to ensure that each retained reasoning step is supported by the scene evidence.

SAVU-Diag is derived from these verified reasoning instances following the procedure described in Sec.~\ref{app:savu_diag_construction}.

\paragraph{Expert Verification.}
After annotation, we conduct a final independent verification stage to assess whether each retained question is clear and solvable from the provided evidence. We recruit three domain experts working in spatial audio or related audio research. The experts independently answer each question. A question is retained only if at least two of the three experts select the intended answer.

This stage provides a final check on evidence sufficiency and removes questions that remain ambiguous after annotation. The final benchmark contains 1,066 SAVU-Bench questions and 611 SAVU-Diag diagnostic questions.

\begin{figure}[!htbp]
    \centering
    \includegraphics[width=0.70\linewidth]
    {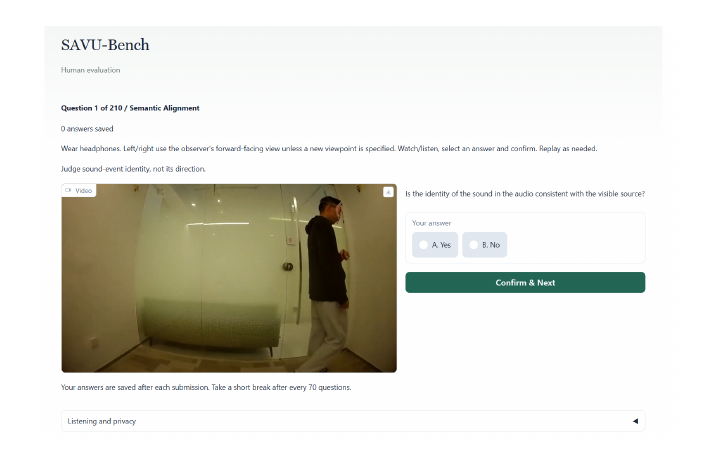}
    \caption{Example of the human evaluation platform.
    }
    \label{fig:human_evaluation}
\end{figure}

\paragraph{Human Evaluation.}
For human evaluation, we randomly sample 30 questions from each of the seven evaluation tasks and evaluate them using a dedicated human-evaluation interface, as shown in Fig.~\ref{fig:human_evaluation}. For human listening, the FOA recordings are rendered into binaural audio using HRTFs and presented together with the same visual inputs used for model evaluation.

We recruit ten participants who are disjoint from the annotators involved in benchmark construction. To cover different levels of domain familiarity, three participants have experience in spatial-audio research, four work in other speech- or audio-related areas, and three have no speech- or audio-related research background. Each participant independently answers all sampled questions once. Human performance for each task group is computed by first calculating the accuracy of each participant and then averaging the accuracies across all ten participants. This protocol provides a common human reference while avoiding overlap between benchmark construction and evaluation.

\end{document}